\documentclass[aps,pre,showpacs,amsmath,amsfonts,amssymb,superscriptaddress,notitlepage,showkeys]{revtex4-1} 

\usepackage[dvips]{graphicx}
\usepackage{bbm}
\usepackage{xcolor}
\usepackage[colorlinks,linkcolor=blue,anchorcolor=blue,citecolor=blue,filecolor=blue,menucolor=blue,urlcolor=blue]{hyperref}
\usepackage{breakurl}

\newcommand{\beq}{\begin{equation}} 
\newcommand{\eeq}{\end{equation}}
\newcommand{\bea}{\begin{eqnarray}} 
\newcommand{\eea}{\end{eqnarray}}

\input epsf 

\begin{document} 

\title{Critical energy density of O$(n)$ models in $d=3$}

\author{Rachele Nerattini}
\email{rachele.nerattini@fi.infn.it}
\affiliation{Dipartimento di Fisica e Astronomia and Centro per lo Studio
delle Dinamiche Complesse (CSDC), Universit\`a di
Firenze, and Istituto Nazionale di Fisica Nucleare (INFN), Sezione di
Firenze, via G.\ Sansone 1, I-50019 Sesto Fiorentino (FI), Italy}

\author{Andrea Trombettoni}
\email{andreatr@sissa.it}
\affiliation{CNR-IOM DEMOCRITOS Simulation Center, via Bonomea 265, I-34136 Trieste, Italy}
\affiliation{SISSA and Istituto Nazionale di Fisica Nucleare (INFN), Sezione di Trieste, via Bonomea 265, I-34136 Trieste, Italy}

\author{Lapo Casetti}
\email{lapo.casetti@unifi.it}
\affiliation{Dipartimento di Fisica e Astronomia and Centro per lo Studio
delle Dinamiche Complesse (CSDC), Universit\`a di
Firenze, and Istituto Nazionale di Fisica Nucleare (INFN), Sezione di
Firenze, via G.\ Sansone 1, I-50019 Sesto Fiorentino (FI), Italy}

\date{\today}

\begin{abstract}
A relation between O$(n)$ models and Ising models has been 
recently conjectured 
[L.\ Casetti, C.\ Nardini, and R.\ Nerattini, Phys.\ Rev.\ Lett.\ {\bf 106}, 
057208 (2011)]. Such a relation, inspired by an energy landscape analysis, 
implies that the microcanonical density of states of an O$(n)$ spin model 
on a lattice can be effectively approximated in terms of the density 
of states of an Ising model defined on the same lattice and 
with the same interactions. Were this relation exact, 
it would imply that the critical energy densities of all the 
O$(n)$ models (i.e., the average values per spin of the O$(n)$ 
Hamiltonians at their respective critical temperatures) 
should be equal to that of the corresponding Ising model; 
it is therefore worth investigating how different the critical 
energies are and how this difference depends on $n$.
 
We compare the critical energy densities of 
O$(n)$ models in three dimensions in some specific cases: 
the O$(1)$ or Ising model, the O$(2)$ or $XY$ model, 
the O$(3)$ or Heisenberg model, the O$(4)$ model and 
the O$(\infty)$ or spherical model, all defined on regular cubic 
lattices and with ferromagnetic nearest-neighbor interactions. 
The values of the critical energy density in the $n=2$, 
$n=3$, and $n=4$ cases are derived through a finite-size 
scaling analysis of data produced by means of Monte Carlo 
simulations on lattices with up to $128^3$ sites. 
For $n=2$ and $n=3$ the accuracy of previously known results 
has been improved. We also derive an interpolation formula showing 
that the difference between the critical energy densities of O$(n)$ 
models and that of the Ising model is smaller than $1\%$ if $n<8$ 
and never exceeds $3\%$ for any $n$.
\end{abstract}


\maketitle
\section{Introduction}
Simple models are important tools in theoretical physics, and 
especially in statistical mechanics, where O$(n)$ Hamiltonians are often used  
to describe in highly simplified, yet significant models  
realistic interactions between particles or spins. Finding links or 
relations between different simple and paradigmatic 
models often results in a deeper understanding of the model themselves 
and of the physics they describe: from this point of view it is highly 
desirable to individuate and characterize exact (or even approximate) 
properties and quantities shared by them.

In \cite{prl2011} a relation between the microcanonical densities 
of states of continuous and discrete spin models was conjectured, 
and further discussed in \cite{jstat2012,analyticalpaper}. 
It was suggested that the density of states of an O$(n)$ 
classical spin model on a given lattice can be approximated 
in terms of the density of states of the corresponding Ising model. 
By ``corresponding'' Ising model we mean an Ising model defined 
on the same lattice and with the same interactions. 
Such a relation was inspired by an energy landscape approach 
\cite{Wales:book} to the microcanonical thermodynamics of these models, 
the key observation being that all the configurations of an Ising model 
on a lattice are stationary points of an O$(n)$ model Hamiltonian 
defined on the same lattice with the same interactions, for any $n$. 
The relation between the densities of states can be written as 
\begin{equation}\label{omega_appr}
\omega^{(n)}(\varepsilon) \approx  \omega^{(1)}(\varepsilon) \,
g^{(n)}(\varepsilon)\, ,
\end{equation}
where $\varepsilon$ is the energy density of the system, 
i.e., $\varepsilon = E/N$ with $E$ and $N$ denoting the 
total energy and the number of spins, respectively; furthermore 
$\omega^{(n)}$ is the density of states of the O$(n)$ model, 
$\omega^{(1)}$ the density of states of the corresponding Ising model 
and $g^{(n)}$ is a function representing the volume of a neighborhood 
of the Ising configuration in the phase space of the O$(n)$ model. 
The function $g^{(n)}$ is typically unknown. However, since it comes from 
local integrals over a neighborhood of the phase space, one expects 
it is regular. Eq.\ (\ref{omega_appr}) is an approximate one 
and the approximations involved are not easily controlled in general 
\footnote{The relation \eqref{omega_appr} cannot be exact, at least in the form proposed in \protect\cite{prl2011}, because it would imply wrong ---and $n$-independent--- 
values of the critical exponent $\alpha$. 
Nevertheless Eq.\ (\ref{omega_appr}) yields the correct sign of $\alpha$, 
that is, correctly predicts a cusp in the specific heat at criticality 
and not a divergence: see Refs.\ \cite{prl2011} and especially 
\cite{jstat2012} for a more complete discussion on the problem.}. 
However, as discussed in \cite{prl2011}, were it exact there would 
be a very interesting consequence: the critical energy 
densities $\varepsilon^{(n)}_{c}$ of the phase transitions 
of all the O$(n)$ models on a given lattice would be the same and 
equal to $\varepsilon^{(1)}_{c}$, that is to the critical 
energy density of the corresponding Ising model. 

Rather surprisingly, according to available analytical and 
numerical calculations the critical energy densities are indeed 
very close to each other whenever a phase transition 
is known to take place, at least for ferromagnetic models 
on $d$-dimensional hypercubic lattices. More precisely, 
the critical energy densities are the same and equal to 
the Ising one for all the O$(n)$ models with long-range mean-field 
interactions as shown by the exact solution 
\cite{CampaGiansantiMoroni:jpa2003}, 
and the same happens for all the O$(n)$ models on a one-dimensional 
lattice with nearest-neighbor interactions. 
Making use of the microcanonical solutions of the models, 
an expression analogous to (\ref{omega_appr}) 
can be exactly computed for the mean-field and for the 
one-dimensional nearest-neighbors $XY$ models ($n=2$) 
\cite{jstat2012}: such expression 
implies the equality of the critical energies 
in the limit $\varepsilon\rightarrow\varepsilon^{(n)}_{c}$. 
Hence the equality of the critical energies is rooted in the 
expression (\ref{omega_appr}) for the density of states.

In $d=2$ the critical energies of the ferromagnetic transition of the 
Ising model and of the Bere\v{z}inskii-Kosterlitz-Thouless (BKT) transition of 
the $XY$ model are only slightly different, 
the difference being about 2\% (see Ref.\ \cite{prl2011} 
and references therein). The thermodynamics of 
the two-dimensional $XY$ model has been analytically studied in 
\cite{analyticalpaper} assuming Eq.\ (\ref{omega_appr}) 
as an ansatz on the form of its density of states 
and then computing $g^{(2)}$ with suitable approximations. 
The results were compared with numerical simulations and a very good 
agreement was found in almost all the energy density range. 
This confirms the soundness of the hypotheses behind Eq.\  (\ref{omega_appr}) 
also in the two-dimensional case. It is also worth noticing 
that despite the difference in the nature of the 
Ising and of the BKT transitions in $d=2$, 
the two-dimensional Ising and $XY$ models share a ``weak universality'': 
indeed, the critical exponent ratio $\beta/\nu$ and the exponent 
$\delta$ are equal in the two cases \cite{Archambault_etal:jpa1997}. 
It is tempting to think that energy landscape arguments like 
those discussed above may explain such a relation 
between the features of phase transitions so different from each other. 

The very different nature, due to the Mermin-Wagner theorem, 
of the Ising and BKT phase transitions in two dimensions  
together with the fact that the comparison is between an exact 
result for $\varepsilon^{(1)}_{c}$ (for the Ising model) 
and numerical results for $\varepsilon^{(2)}_{c}$ (for the XY model) 
prevents the two-dimensional case from being 
a good test case to quantify the accuracy of the prediction 
on the equality of critical energy densities. From this point of view 
the O$(n)$ model in three dimensions ($d=3$) provides a very promising 
and clear-cut case study to test the equality of the critical energy 
densities since a phase transition occurs for all $n$ and in all cases a local 
order parameter becomes non-vanishing at a finite critical temperature.  
For nearest-neighbor interacting O$(n)$ models in 
$d = 3$ the comparison has to be based on the outcomes 
of numerical simulations or on approximate 
methods, since no exact solution (in particular for the critical energy) 
exists even for the Ising case. Although typically overlooked, 
results reported in the literature clearly show that the critical energies 
measured for three-dimensional O$(n)$ spin systems with $n = 1$, $2$ and $3$ 
are almost consistent: see \cite{prl2011} for a discussion on this point and 
\cite{BradyMoreira:prb1993,GottlobHasenbusch:physicaa1993,BrownCiftan:prb2006} 
for the critical values of the energy densities for $n=1$, 
$n=2$ and $n=3$, respectively. 

Inspired by these results, the aim of this paper is 
to quantify the difference between the critical energy densities 
of nearest-neighbor O$(n)$ models defined on regular 
cubic lattices in $d=3$ and to study the dependence on $n$ of 
the O$(n)$ critical energy densities. 
This study also entails an assessment of the accuracy of the prediction 
of equal critical energy densities following from Eq.\ \eqref{omega_appr}.

As shown in the following Sections, the already existing 
numerical estimates of the critical energy densities for 
three-dimensional O$(n)$ models with $n=2$ and $3$ will be improved; 
in the case $n=4$ we obtain a result having the 
same accuracy of, and in good agreement with, 
a very recent one given in \cite{EngelsKarsch:prd2012}. 
Using these results together with the exact 
result for the critical energy density of the $n=\infty$ 
model (i.e., the spherical model \cite{Stanley:physrev1968}) and 
with the first term of the 
$1/n$ expansion \cite{CampostriniEtAl:npb1996},  
an interpolation formula for the critical energy densities 
$\varepsilon^{(n)}_{c}$ will be derived, valid in 
the whole range $n=1,2,\ldots,\infty$. 
It will turn out that the difference between the critical 
energy densities of the O$(n)$ models and that 
of the corresponding Ising model is smaller than $1\%$ for 
O$(n)$ models with $n<8$ and never exceeds $3\%$.

The paper is organized as follows: In Sec.\ \ref{Onmodels} 
the definition of O$(n)$ models is recalled and the 
notation used in the next Section 
introduced. Assuming the critical energy density of the Ising model 
in three dimensions known with enough accuracy 
\cite{HasenbuschPinn:jphysa1998}, in Sec.\ \ref{SecFSS} we 
estimate the critical energy densities 
of the O$(2)$, O$(3)$ and O$(4)$ models in $d=3$ via a finite-size 
scaling (FSS) analysis whose basic relations are presented in 
Sec.\ \ref{SecFSS}. In Sec.\ \ref{numericalSphericalModel} 
the spherical model in $d=3$ is discussed since its 
thermodynamics is 
equivalent to the one of an O$(n)$ model in the $n\rightarrow\infty$ limit. 
The spherical model can be solved analytically in any spatial dimension $d$ 
and, in particular, in $d=3$: hence it provides the value of 
$\varepsilon^{(\infty)}_{c}$. In Sec.\ \ref{Sec_NumericalTest} 
a careful comparison between the critical values of the energy 
densities of the above mentioned models is performed and an interpolation 
formula for $\varepsilon^{(n)}_{c}$ defined. 
Some conclusions are drawn in Sec.\ \ref{ConclusionsNumericalTest}. 

\section{O$(n)$ spin models}\label{Onmodels}
In the following we are going to consider classical O$(n)$ spin 
models defined on a regular cubic lattice in $d=3$ and with 
periodic boundary conditions. To each lattice site $i$ an 
$n$-component classical spin vector $\mathbf{S}_i = (S_i^1,\ldots,S_i^n)$ 
of unit length is assigned. The energy of the model is given by the Hamiltonian
\begin{equation}\label{H-On}
H^{(n)} = - J \sum_{\langle i,j \rangle} \mathbf{S}_i \cdot \mathbf{S}_j= 
- J \sum_{\langle i,j \rangle} \sum_{a = 1}^n S^a_i S^a_j\, ,
\end{equation}
where the angular brackets denote a sum over all distinct pairs of 
nearest-neighbor lattice sites. 
The exchange coupling $J$ will be assumed positive, 
resulting in ferromagnetic interactions. 
The Hamiltonian (\ref{H-On}) is globally invariant under the $O(n)$ group. 

In the special cases $n=1$, $n=2$, and $n=3$, one obtains 
the Ising, $XY$, and Heisenberg models, respectively. 
The case $n=1$ is even more special because O$(1) \equiv \mathbb{Z}_2$ 
is a discrete symmetry group. 
In this special case the Hamiltonian (\ref{H-On}) becomes the Ising Hamiltonian 
\begin{equation}
H^{(1)} = - J \sum_{i,j=1}^N \sigma_i \sigma_j~\, ,
\label{H_1}
\end{equation}
where $\sigma_i = \pm 1$ $\forall i$. In all the other cases $n \geq 2$ the 
O$(n)$ group is continuous. 
Without loss of generality we shall set $J=1$ in the following (and 
$k_B=1$). 

The energy density $\varepsilon = H^{(n)}/N$ 
lies in the energy range $[-d,d]$ where $d$ is the lattice dimension. 
In $d=3$ and for any $n$ the models exhibit a phase transitions 
at $\varepsilon=\varepsilon^{(n)}_{c}$ from a paramagnetic phase, 
for $\varepsilon>\varepsilon^{(n)}_{c}$, to a ferromagnetic phase, 
for $\varepsilon<\varepsilon^{(n)}_{c}$, with a spontaneous breaking 
of the O$(n)$ symmetry. The models are not exactly solvable and 
estimates of critical temperatures, critical exponents and other quantities 
at criticality have been mainly derived by means of 
numerical simulations, see e.g.\ 
\cite{BradyMoreira:prb1993,GottlobHasenbusch:physicaa1993,BrownCiftan:prb2006}.

\section{Determination of the critical energy 
densities}\label{SecNumericalIntro}
The aim of this work is to answer the following question: 
what is the difference between the critical value $\varepsilon^{(n)}_{c}$ 
of the energy density of the O$(n)$ model (\ref{H-On}) and 
the critical value $\varepsilon^{(1)}_{c}$ of the energy density 
of the Ising model (\ref{H_1})? And how does it depend on $n\in[2,\infty]$? 

Some preliminary observations are necessary. 
As mentioned in the Introduction, three-dimensional O$(n)$ 
models are not exactly solvable 
\footnote{But in the case $n\rightarrow\infty$, that will be discussed 
in Sec.\ \ref{numericalSphericalModel}.}
and the value of thermodynamic functions at criticality  
is typically estimated numerically. 

Most numerical simulations have been limited so far mostly to small $n$: 
see e.g.\ \cite{BradyMoreira:prb1993,GottlobHasenbusch:physicaa1993,BrownCiftan:prb2006,EngelsKarsch:prd2012} for $n=1$, $2$, $3$ and $4$, respectively. This is clearly understandable since these are the most relevant cases for physical applications \cite{CampostriniEtAl:npb1996}. On the other hand, different approaches like $1/n$ and strong-coupling expansions have been used for large $n$, see Ref.\  \cite{CampostriniEtAl:npb1996}.
The common feature of these studies is that they have been performed in 
the canonical ensemble. Hence, especially before the suggestion 
that critical energy densities might be very close or even equal 
\cite{prl2011}, 
an accurate evaluation of the critical energy densities 
$\varepsilon^{(n)}_{c}$ was out of the scope of the works, 
and the computation of $\varepsilon^{(n)}_{c}$ was usually 
a byproduct of a more general task possibly focused on the 
determination of other parameters, 
such as the critical temperatures $T^{(n)}_{c}$ or the critical exponents 
or the free energies at the critical point. 
In the following we shall use Monte Carlo simulations and 
FSS to determine improved estimates of $\varepsilon^{(n)}_{c}$ 
for $n=2$ and $3$, our estimate of $\varepsilon^{(4)}_{c}$ 
being as accurate as the most recent in the literature 
\cite{EngelsKarsch:prd2012}. The case $n=1$ has already been studied 
with high accuracy by Hasenbusch and Pinn in 
\cite{HasenbuschPinn:jphysa1998} and we 
will simply recall their results in Sec.\ \ref{numericalO1model}. 

The FSS analyses rely on numerical data computed by means 
of canonical Monte Carlo simulations using 
the optimized cluster algorithm {\tt{spinmc}} 
for classical O$(n)$ spin models provided by the ALPS project \cite{ALPS}. 
Most of the simulations have been performed on the PLX 
machine at the CINECA in Casalecchio di Reno (Bologna, Italy). 
A small subset of the simulations has been performed with the same 
{\tt{spinmc}} algorithm on the PC-farm of the Dipartimento 
di Fisica e Astronomia of the Universit\`{a} di Firenze, Italy. 
We typically used $5\times 10^6$ Monte Carlo sweeps (MCS) 
plus $5\times 10^5$ MCS of thermalization for 
the simulations of the O$(2)$ model and $10^7$ MCS plus 
$2.5\times 10^6$ MCS of thermalization for the simulations of 
the O$(3)$ and of the O$(4)$ model. The total cluster CPU 
time spent on PLX for the simulations has been more than $40000$ hours. 

For each O$(n)$ model, the simulations have been performed at the 
value of the critical temperature $T^{(n)}_{c}$ 
given in the literature with an uncertainty $\varDelta T^{(n)}_{c}$. 
This quantity has to be taken into account 
in the computation of the uncertainty $\varDelta \varepsilon^{(n)}_{c}$ 
associated to the estimate of $\varepsilon^{(n)}_{c}$ and 
the uncertainty propagation procedure needs 
the evaluation of the critical value of the specific heat. 
For this reason, in the Monte Carlo simulations, 
besides collecting the values of the energy densities, 
we also computed the specific heat. 
The FSS procedure and the uncertainty propagation procedure will 
be discussed in the following section.

\subsection{Finite-size scaling analysis}\label{SecFSS}
Let us denote by $\varepsilon^{(n)}_{c}(L)$ and $c^{(n)}(L)$ the 
critical values of the energy density and of the specific 
heat, respectively, of an O$(n)$ model defined on a regular cubic lattice 
of edge $L=\sqrt[3]{N}$. 
The relation between $\varepsilon^{(n)}_{c}(L)$ and 
$\varepsilon^{(n)}_{c}(\infty) \equiv \varepsilon^{(n)}_{c}$ 
is given by the FSS equation 
\begin{equation}\label{energy_FSS}
\varepsilon^{(n)}_{c}(L) = \varepsilon^{(n)}_{c}+\varepsilon_{n}\;L^\frac{\alpha_{n}-1}{\nu_{n}}\, :
\end{equation}
in the following we use the notation 
\begin{equation}\label{definition_Dn}
D_{n}=\frac{\alpha_{n}-1}{\nu_{n}}\, .
\end{equation}
An analogous expression holds for the specific heat, and it is given by
\begin{equation}\label{cv_FSS}
c^{(n)}(L)=c^{(n)}_{c}+c_{n}\;L^\frac{\alpha_{n}}{\nu_{n}}\, ,
\end{equation}
where $c^{(n)}_{c} \equiv c^{(n)}_{c}(\infty)$ denotes the 
critical value of the specific heat in the thermodynamic limit. 
In Eqs.\ (\ref{energy_FSS}) and (\ref{cv_FSS}), $\varepsilon_{n}$ and $c_{n}$ are model dependent fit parameters, 
while $\alpha_{n}$ and $\nu_{n}$ are the specific heat and 
the correlation length critical exponents, respectively. 
We do not discuss here the derivation of 
Eqs.\ (\ref{energy_FSS}) and (\ref{cv_FSS}), referring 
the reader to the existing literature for an in-depth analysis on the subject, 
see e.g.\ \cite{Fisher:rmp1974,Brezin:jphys1982,Stanley:rmp1999} 
for reviews and 
\cite{SchultkaManousakis:prb1995} for an explicit derivation of 
Eqs.\ (\ref{energy_FSS}) and (\ref{cv_FSS}) in the case $n=2$.

For each O$(n)$ model, the estimate of the critical energy 
density $\varepsilon^{(n)}_{c}\pm\varDelta \varepsilon^{(n),stat}_{c}$ 
can be determined with a fit of the Monte Carlo data 
$\varepsilon^{(n)}_{c}(L)$ according to Eq.\ (\ref{energy_FSS}); 
here and in the following $\varDelta \varepsilon^{(n),stat}_{c}$ will denote 
the statistical uncertainty on $\varepsilon^{(n)}_{c}$ 
due to the fitting procedure. 

Since our purpose is to compare the values of $\varepsilon^{(n)}_{c}$ 
for different $n$, any source of error in the 
determination of $\varDelta \varepsilon^{(n)}_{c}$ 
has to be considered separately.  
The fact that the energy data $\varepsilon^{(n)}_{c}(L)$ 
are computed with Monte Carlo simulations performed at $T^{(n)}_{c}$ 
becomes important. Indeed, the critical temperatures $T^{(n)}_{c}$ of O$(n)$ 
models are provided in the literature with an uncertainty 
$\varDelta T^{(n)}_{c}$ whose effect in the determination 
of $\varDelta \varepsilon^{(n)}_{c}$ has to be checked with special care. 
As a matter of fact, $\varDelta T^{(n)}_{c}$ can be seen as the analogous 
of a systematic source of error in an experimental setting; 
we will then denote by $\varDelta \varepsilon^{(n),syst}_{c}$ 
its contribution to $\varDelta \varepsilon^{(n)}_{c}$. 
The two contributions $\varDelta \varepsilon^{(n),stat}_{c}$ and 
$\varDelta \varepsilon^{(n),syst}_{c}$ to the uncertainty 
$\varDelta\varepsilon^{(n)}_{c}$ of $\varepsilon^{(n)}_{c}$ 
will be discussed separately in the following, 
and the final estimate of $\varepsilon^{(n)}_{c}$ will be given in the form  
\begin{equation}\label{final-e-estimation}
\varepsilon^{(n)}_{c}\pm \varDelta \varepsilon^{(n)}_{c} \equiv 
\varepsilon^{(n)}_{c}\pm \varDelta \varepsilon^{(n),stat}_{c} 
\pm \varDelta \varepsilon^{(n),syst}_{c}\, .
\end{equation}

The systematic uncertainty $\varDelta \varepsilon^{(n),syst}_{c}$ can be 
estimated with two different methods. In both cases the critical 
value $c^{(n)}_{c}$ of the specific heat is necessary and 
will be computed with a fit 
\footnote{For $c^{(n)}_{c}$ only the statistical error 
$\varDelta c^{(n),stat}_{c}$ will be computed since this quantity is only 
used for the computation of $\varDelta \varepsilon^{(n),syst}_{c}$.} 
of the Monte Carlo data $c^{(n)}_{c}(L)$ according to Eq.\ (\ref{cv_FSS}). 
The two methods we used to compute $\varDelta \varepsilon^{(n),syst}_{c}$ 
are the following:
\begin{itemize}
\item \textit{Method 1.}
\begin{equation}\label{second-energy}
\varDelta \bar{\varepsilon}^{(n),syst}_{c} = |\varepsilon^{(n)}_{c} - \bar{\varepsilon}^{(n)}_{+} | = 
|\varepsilon^{(n)}_{c} - \bar{\varepsilon}^{(n)}_{-} |\, :
\end{equation}
$\bar{\varepsilon}^{(n)}_{\pm}$ denote the energy densities at $T^{(n)}_{\pm}=T^{(n)}_{c}\pm\varDelta T^{(n)}_{c}$, 
computed with a first order Taylor expansion around $\varepsilon^{(n)}_{c}$; that is,
\begin{equation}\label{Taylor-espansion}
\begin{split}
\bar{\varepsilon}^{(n)}_{\pm}= &\varepsilon^{(n)}_{c}\Big|_{T=T^{(n)}_{c}} + 
\frac{d\varepsilon}{dT}\Big|_{T=T^{(n)}_{c}}\;\left[\left(T^{(n)}_{c}\pm\varDelta T^{(n)}_{c} \right)- T^{(n)}_{c}\right]=\\
 =& \varepsilon^{(n)}_{c}\pm c^{(n)}_{c}\;\varDelta T^{(n)}_{c}\, .
\end{split}
\end{equation}
\item \textit{Method 2.}
\begin{equation}
\varDelta \tilde{\varepsilon}^{(n),syst}_{c}=
\cdot^{|\varepsilon^{(n)}_{c} - \tilde{\varepsilon}^{(n)}_{+}|}_{|\varepsilon^{(n)}_{c} - \tilde{\varepsilon}^{(n)}_{-}|}\, ,
\end{equation}
with $\tilde{\varepsilon}^{(n)}_{\pm}$ 
denoting again the energy density values at $T^{(n)}_{\pm}$;
at variance $\bar{\varepsilon}^{(n)}_{\pm}$, $\tilde{\varepsilon}^{(n)}_{\pm}$ 
are computed with a 
fit of the energy density data $\tilde{\varepsilon}^{(n)}_{\pm}(L)$ at $T^{(n)}_{\pm}$. The values of 
$\tilde{\varepsilon}^{(n)}_{\pm}(L)$ are computed in part  
with a first order Taylor expansion of the 
numerical data for $\varepsilon^{(n)}_{c}(L)$ through the relation
\begin{equation}\label{Taylor-espansion-2}
\begin{split}
\tilde{\varepsilon}^{(n)}_{\pm}(L) &= \varepsilon^{(n)}(L) \Big|_{T=T^{(n)}_{c}}+ c^{(n)}_{c}(L)\Big|_{T=T^{(n)}_{c}}
\left[\left(T^{(n)}_{c}\pm\varDelta T^{(n)}_{c}\right)-T^{(n)}_{c}\right]=\\
&=\varepsilon^{(n)}(L)\pm c^{(n)}_{c}(L)\;\varDelta T^{(n)}_{c}\, ,
\end{split}
\end{equation}
and in part ---namely for $L=32$, $64$ and $128$--- numerically 
by performing Monte Carlo simulations of the systems at $T^{(n)}_{\pm}$ 
(the two procedures give results for $\tilde{\varepsilon}^{(n)}_{\pm}(L)$ 
in excellent agreement).

In the end, the fitting procedure is applied according to the relation 
\footnote{Notice that Eqs.\ (\ref{energy_FSS}) and 
Eq.\ (\ref{energy_FSS_out}) hold for $T=T^{(n)}_{c}$ - 
however, since $\frac{\varDelta T^{(n)}_{c}}{T^{(n)}_{c}}\sim 10^{-5}$ 
for the models considered, we assume Eq.\ (\ref{energy_FSS_out}) 
valid in the whole range
$T\in\left[T^{(n)}_{c}-\varDelta T^{(n)}_{c}, 
T^{(n)}_{c}+\varDelta T^{(n)}_{c}\right]$.}
\begin{equation}\label{energy_FSS_out}
 \tilde{\varepsilon}^{(n)}_{\pm}(L)=\tilde{\varepsilon}^{(n)}_{\pm} +\varepsilon_{n,\pm} L^{D_{n}}
\end{equation}
with $D_{n}$ given in Eq.\ (\ref{definition_Dn}).  
\end{itemize}
At the end of the analysis, $\varDelta \bar{\varepsilon}^{(n),syst}_{c}$ and 
$\varDelta \tilde{\varepsilon}^{(n),syst}_{c}$ will be compared and one of 
them will be chosen as final estimate of $\varDelta \varepsilon^{(n),syst}_{c}$.

\subsection{$n=1$, the Ising model}\label{numericalO1model}
The derivation of the critical energy density 
$\varepsilon^{(1)}_{c}$ for the three-dimensional 
Ising model can be found in Ref.\ \cite{HasenbuschPinn:jphysa1998}: 
the authors performed a FSS analysis of data 
computed with canonical Monte Carlo simulations of the system, 
considering lattices up to $112^{3}$ spins. The critical coupling 
$\beta^{(1)}_{c} \equiv 1/T^{(1)}_{c}$ reported in 
\cite{HasenbuschPinn:jphysa1998,TalapovBlote:jphysa1996} is 
$\beta^{(1)}_{c}=0.2216544(6)$ [see as well the discussion in 
\cite{Binder:book}, p.\ 265 (Chapter 7), and references therein]. 
The best final estimate of the critical energy density is given by
\begin{equation}\label{Ising_best_energy}
\varepsilon^{(1)}_{c} \pm \varDelta\varepsilon^{(1)}_{c} =-0.99063 \pm 0.00004\, .
\end{equation}
The above result has been computed considering system sizes close 
to the maximum achievable with our tools and represents 
one of the most accurate estimation of $\varepsilon^{(1)}_{c}$ 
available in the literature (see, e.g., \cite{BradyMoreira:prb1993} 
for a comparison). Moreover, the uncertainty $\varDelta \varepsilon^{(1)}_{c}$ in 
Eq.\ (\ref{Ising_best_energy}) has been computed 
combining the statistical and the systematic error as we have 
discussed in the previous Section. 
These facts led us not to repeat the analysis on 
the Ising model and to consider 
Eq.\ (\ref{Ising_best_energy}) as the best final estimation of $\varepsilon^{(1)}_{c}$. A further comment on this point can 
be found in Sec.\ \ref{ConclusionsNumericalTest}.

\subsection{$n=2$, the XY model}\label{numericalO2model}
We performed canonical Monte Carlo simulations of the 
$XY$ model defined on regular 
cubic lattices with edges $L=32,40,50,64,80,100$ and $128$. 
The simulations have been performed at a temperature 
$T = 2.201673$ according to the critical 
value of the temperature $T^{(2)}_{c}=2.201673(97)$ 
reported in \cite{GottlobHasenbusch:physicaa1993}. The values for 
$\varepsilon^{(2)}_{c}(L)$ and $c^{(2)}_{c}(L)$ obtained from the 
simulations are reported in Table \ref{table:xy_data_ec}: 
in parentheses are the statistical errors. 

\begin{table}[ht]
\caption{{Monte Carlo results for the energy 
density $\varepsilon^{(2)}_{c}(L)$ and for the specific 
heat $c^{(2)}_{c}(L)$ at the critical temperature $T^{(2)}_{c}=2.201673$.}}
\centering
\begin{tabular}{|c |c |c|}
\hline
$L$ & $\varepsilon^{(2)}_{c}(L)$ & $c^{(2)}_{c}(L)$  \\
[0.5ex]
\hline
\hline 
32 & -0.9982(3) & 2.611(31) \\ 

40 & -0.99589(12) & 2.709(18) \\

50 & -0.99382(9) & 2.825(24)\\

64 & -0.99233(14) & 2.923(59) \\

80 & -0.99137(6) & 3.074(34) \\

100 & -0.99067(4) & 3.199(38) \\

128 & -0.99020(4) & 3.282(54) \\ [1ex] 
\hline
\end{tabular}
\label{table:xy_data_ec}
\end{table}

We fitted the energy density data reported in Table \ref{table:xy_data_ec} 
according to the relation 
(\ref{energy_FSS}) considering different choices for the critical exponents. 
In particular we chose: 
(\textit{i}) the experimental values $\nu_2 = 0.6705(6)$ and 
$\alpha_2 = -0.0115(18)$ as 
reported in \cite{GoldnerAhlers:prb1992}; (\textit{ii}) 
$\nu_2=0.662(7)$ obtained in \cite{GottlobHasenbusch:physicaa1993} 
at the same critical value of the 
temperature as in our case and $\alpha_2=-0.014(21)$ 
as derived from the scaling relation $\alpha_2=2-d\nu_2$ with $d=3$; 
(\textit{iii}) $\nu_2=0.6723(3)$ obtained in 
\cite{HasenbuschTorok:jphysa1999} 
with a high statistics simulation performed at a 
slightly different value of the temperature and $\alpha_2=-0.017(3)$ 
as derived from the scaling relation $\alpha=2-d\nu$ with $d=3$; 
(\textit{iv}) $\alpha_{2}/\nu_{2}=-0.0258(75)$ and $1/\nu_{2}=1.487(81)$ as 
obtained in \cite{SchultkaManousakis:prb1995} 
with a similar analysis. 
The results of the fits for $\varepsilon^{(2)}_{c}$ and 
for the fitting parameter $\varepsilon_{2}$ are reported 
in Table \ref{table:xy_fit_results}. We also performed a four-parameters fit 
considering $\alpha_2$, $\nu_2$, $\varepsilon^{(2)}_{c}$ and $\varepsilon_{2}$ 
as free parameters. However, no meaningful results 
could be extracted from the fit, the relative error on the 
parameters being larger than $100\%$ on the critical exponents 
(data not shown). 

\begin{table}[ht]
\caption{{Fitting values of the parameters $\varepsilon^{(2)}_{c}$ and $\varepsilon_{2}$ 
entering expression (\ref{energy_FSS}).}}
\centering
\begin{tabular}{|c| c| c| c|}
\hline
Fitting parameters & $\nu_2$ and $\alpha_2$ & results & $\chi^2/\text{d.o.f.}$ \\
[0.5ex]
\hline
\hline
 & {$\nu_2=0.6705$}     & {$\varepsilon^{(2)}_{c}=-0.98900(3)$} &\\
 [-1ex]
\raisebox{1.5ex}{{$\varepsilon^{(2)}_{c},\;\varepsilon_{2}$}}       & 
{$\alpha_2=-0.0115$} &{$\varepsilon_{2}=-1.77(2)$}       &
\raisebox{1.5ex}{{$0.60$}} \\
\hline 

 &  {$\nu_2=0.662$}     &  {$\varepsilon^{(2)}_{c}=-0.98904(3)$} &\\[-1ex]
\raisebox{1.5ex}{ {$\varepsilon^{(2)}_{c},\;\varepsilon_{2}$}}       & 
 {$\alpha_2=-0.014$} &  {$\varepsilon_{2}=-1.92(2)$}       &
\raisebox{1.5ex}{ {$0.57$}} \\
\hline 

 &  {$\nu_2=0.6723$}     &  {$\varepsilon^{(2)}_{c}=-0.98901(3)$} &\\[-1ex]
\raisebox{1.5ex}{ {$\varepsilon^{(2)}_{c},\;\varepsilon_{2}$}}       & 
 {$\alpha_2=-0.017$} &  {$\varepsilon_{2}=-1.79(2)$}       &
\raisebox{1.5ex}{ {$0.59$}} \\
\hline 

 &  {$\alpha_2/\nu_2=-0.0258$}     &  {$\varepsilon^{(2)}_{c}=-0.98901(3)$} &\\[-1ex]
\raisebox{1.5ex}{ {$\varepsilon^{(2)}_{c},\;\varepsilon_{2}$}}       & 
 {$1/\nu_2=1.487$} &  {$\varepsilon_{2}=-1.79(2)$}       &
\raisebox{1.5ex}{ {$0.59$}} \\
\hline 
\end{tabular}
\label{table:xy_fit_results}
\end{table}

\begin{figure}
\centering
\includegraphics[width=10cm]{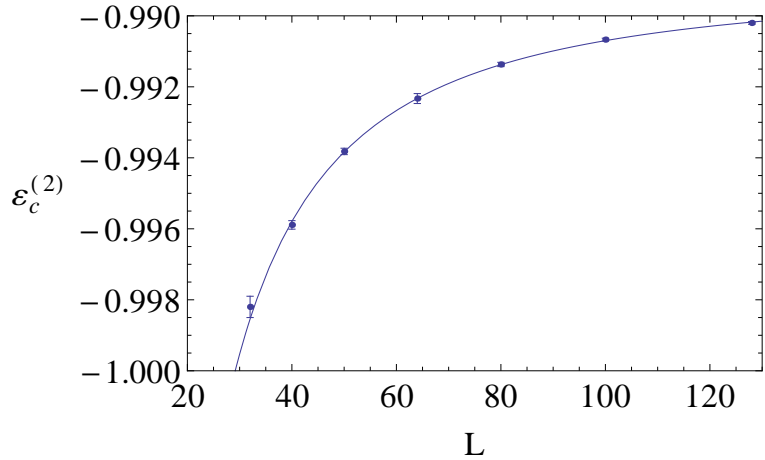} 
\caption{ {Energy density $\varepsilon^{(2)}_{c}$ at the critical temperature $T^{(2)}_{c}=2.201673$ 
as a function of $L$. The solid curve is the fit to (\ref{XY_energy_I_best_result}) with $\nu_2=0.662$ and $\alpha_2=-0.014$.}} \label{xy_energy_plot}
\end{figure}

All the results reported in 
Table \ref{table:xy_fit_results} have a 
$\chi^2/\text{d.o.f.}\simeq 0.6$ and all the values of the 
critical energy densities $\varepsilon^{(2)}_{c}$ are consistent with each other. 
This fact implies that $\varepsilon^{(2)}_{c}$ is rather 
insensitive to the choice of the critical exponents  $\nu_2$ and $\alpha_2$ 
(and so to the values of the critical 
temperatures at which they have been computed). 
Anyway, as best estimate of the fitting parameters we chose:
\begin{equation}\label{XY_energy_I_best_result}
\begin{split}
\varepsilon^{(2)}_{c} \pm \varDelta \varepsilon^{(2),stat}_{c} &=-0.98904\;\pm\;0.00003\, , \\
\varepsilon_{2} &=-1.92\;\pm\;0.02
\end{split}
\end{equation}
reported in the second row of Table \ref{table:xy_fit_results}. 
These values correspond to a choice of the critical 
exponents given by $\nu_2=0.662$ and $\alpha_2=-0.014$ as derived 
in \cite{GottlobHasenbusch:physicaa1993} (second raw 
of Table \ref{table:xy_fit_results}) assuming 
the same value of $T^{(2)}_{c}$ as in our case. 
The curve $\varepsilon^{(2)}_{c}(L)$ given by 
Eq.\ (\ref{energy_FSS}) for $n=2$ and with the values of 
$\varepsilon^{(2)}_{c}$ and $\varepsilon_{2}$ as in 
Eq.\ (\ref{XY_energy_I_best_result}) is shown in Fig.\ \ref{xy_energy_plot} 
together with the simulation data. 
$\varepsilon^{(2)}_{c}$ and $\varepsilon_{2}$ in 
Eq.\ (\ref{XY_energy_I_best_result}) are consistent with the values reported 
in \cite{SchultkaManousakis:prb1995}; therein, 
authors found $\varepsilon^{(2)}_{c}=-0.9890(4)$ and 
$\varepsilon_{2}=-1.81(38)$.
It is worth noticing that our result $\varepsilon^{(2)}_{c}=-0.98904(3)$ 
given in Eq.\ (\ref{XY_energy_I_best_result}) has 
one digit of precision more than previous 
results obtained with analogous techniques, 
see e.g.\ \cite{SchultkaManousakis:prb1995}.

We fitted data of $c^{(2)}_{c}(L)$ reported 
in Table \ref{table:xy_data_ec} according to the scaling relation 
given in Eq.\ (\ref{cv_FSS}) and keeping the value of the ratio 
$\alpha_2/\nu_2$ constant and equal to $\alpha_2/\nu_2=-0.02$, 
as given in \cite{GottlobHasenbusch:physicaa1993}. 
The result of the fit is reported in the first row of 
Table \ref{table:xy_cv_fit_results}. 
To check the dependence of the specific heat on 
the value of the ratio $\alpha_2/\nu_2$, we 
also performed the same fit for different values of 
the critical exponents: (\textit{i}) $\alpha_2/\nu_2=-0.0285$ as reported 
in \cite{SchultkaManousakis:prb1995}; 
(\textit{ii}) $\alpha_2/\nu_2=-0.025$ as obtained 
from data in \cite{HasenbuschTorok:jphysa1999}; 
(\textit{iii}) $\alpha_2/\nu_2=-0.0172$ 
as obtained from the experimental values of the critical exponents 
reported in \cite{GoldnerAhlers:prb1992}. 
The results of the fits for $c^{(2)}_{c}$ and $c_{2}$ with these choices 
of the critical exponents are reported in the second, 
third and fourth row of Table 
\ref{table:xy_cv_fit_results}, respectively. 

Although the values of $c^{(2)}_{c}$ 
reported in Table \ref{table:xy_cv_fit_results} are not 
all consistent with each other, 
the results in the first three rows are comparable. 
Moreover, our results assuming $\alpha_2/\nu_2=-0.0285$ 
are in agreement with the results computed 
in \cite{SchultkaManousakis:prb1995} with the same 
choice of the ratio of the critical exponents. 
Indeed, authors found $c^{(2)}_{c}=20.45(66)$ and $c_{2}=-19.61(72)$ 
with a fit based on data derived form Monte Carlo 
simulations at a different value of the critical temperature. 
Interestingly the values of the fitting 
parameters $c^{(2)}_{c}$ and $c_{2}$ are slightly larger when the experimentally 
determined critical exponents $\nu_2=0.6705$ and $\alpha_2=-0.0115$ \cite{GoldnerAhlers:prb1992} are considered, 
see the last row of Table \ref{table:xy_cv_fit_results}. 
This fact was already pointed 
out in \cite{SchultkaManousakis:prb1995} 
where the authors found $c^{(2)}_{c}=30.3\pm1.0$ and $c_{2}=-29.4\pm1.1$ 
for the same choice of the critical exponents. 
These results suggest that the value of $c^{(2)}_{c}$ 
strongly depends on the value 
of the ratio $\alpha_2/\nu_2$. In \cite{SchultkaManousakis:prb1995} 
the authors considered 
lattice sizes up to $L=80$ and suggested that a wider range of lattice sizes 
should be necessary to determine the asymptotic value of $c^{(2)}_{c}$. 
In our analysis we considered lattice sizes up to 
$L=128$, giving $N$ almost $4$ 
times bigger than in \cite{SchultkaManousakis:prb1995}, 
but the discrepancy is still visible. 
Lattice sizes bigger than $128^3$ spins 
may be needed to improve the estimate of $c^{(2)}_{c}$. 
For our purposes, we can consider
\begin{equation}\label{XY_cv_best_result}
\begin{split}
c^{(2)}_{c}\pm \varDelta c^{(2)}_{c} &=28.4\pm0.6\,,\\
c_{2}&=-27.7\pm0.7
\end{split}
\end{equation}
as best final estimates of the fitting parameters. 
These quantities, in fact, derive 
from the fit with $\alpha_2/\nu_2=-0.02$ as obtained 
in \cite{GottlobHasenbusch:physicaa1993} assuming 
the same value of the critical temperature $T^{(2)}_{c}=2.201673$ as in our case. 
We refer the reader to \cite{SchultkaManousakis:prb1995} for a more 
detailed discussion of this problem. 

\begin{table}[ht]
\caption{ {Fitting values of the parameters $c^{(2)}_{c}$ and $c_{2}$ entering expression (\ref{cv_FSS}).}}
\centering
\begin{tabular}{|c| c| c| c|}
\hline
Fitting parameters & $\alpha_2/\nu_2$ & results & $\chi^2/\text{d.o.f.}$ \\
[0.5ex]
\hline
\hline
 &	&  {$c^{(2)}_{c}=28.4\pm0.6$} &\\[-1ex]
\raisebox{1.5ex}{ {$c^{(2)}_{c},\;c_{2}$}}       &
\raisebox{1.5ex}{ {$\alpha_2/\nu_2=-0.02$}}  &  {$c_{2}=-27.7\pm0.7$}       &
\raisebox{1.5ex}{ {$0.2$}} \\
\hline 

& 	&  {$c^{(2)}_{c}=22.7\pm0.5$} &\\[-1ex]
\raisebox{1.5ex}{ {$c^{(2)}_{c},\;c_{2}$}}       & 
\raisebox{1.5ex}{ {$\alpha_2/\nu_2=-0.0258$}} &  {$c_{2}=-21.9\pm0.5$}       &
\raisebox{1.5ex}{ {$0.2$}} \\
\hline 


 & 	&  {$c^{(2)}_{c}=23.3\pm0.5$} &\\[-1ex]
\raisebox{1.5ex}{ {$c^{(2)}_{c},\;c_{2}$}}       & 
\raisebox{1.5ex}{ {$\alpha_2/\nu_2=-0.025$}} &  {$c_{2}=-22.6\pm0.6$}       &
\raisebox{1.5ex}{ {$0.2$}} \\
\hline  

 & 	&  {$c^{(2)}_{c}=32.5\pm0.7$} &\\[-1ex]
\raisebox{1.5ex}{ {$c^{(2)}_{c},\;c_{2}$}}       & 
\raisebox{1.5ex}{ {$\alpha_2/\nu_2=-0.0172$}} &  {$c_{2}=-31.8\pm0.8$}       &
\raisebox{1.5ex}{ {$0.2$}} \\
\hline 

\end{tabular}
\label{table:xy_cv_fit_results}
\end{table}

\begin{figure}
\centering
\includegraphics[scale=0.9]{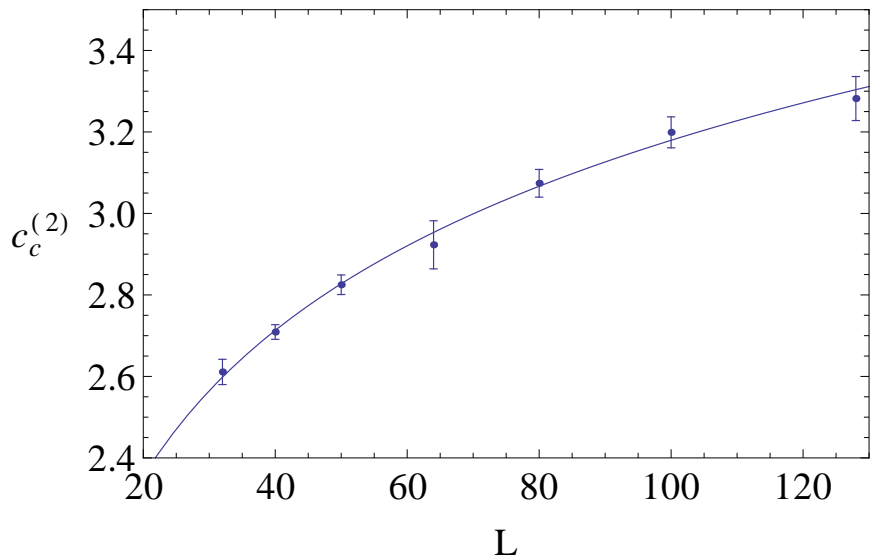} 
\caption{ {Specific heat $c^{(2)}_{c}$ at the critical temperature $T^{(2)}_{c}=2.201673$ 
as a function of $L$. The solid curve represents the fit to (\ref{XY_cv_best_result}) with $\alpha_2/\nu_2=-0.02$.}}
\label{xy_cv_plot}
\end{figure}

The curve $c^{(2)}_{c}(L)$ given by Eq.\ (\ref{cv_FSS}) for $n=2$ 
with $c^{(2)}_{c}$ and $c_{2}$ as in 
Eq.\ (\ref{XY_cv_best_result}) is plotted in Fig.\ \ref{xy_cv_plot} 
together with the simulation data.

In order to evaluate the systematic contribution to the uncertainty, 
$\varDelta \varepsilon^{(2),syst}_{c}$, we applied the two methods 
presented in Sec.\ \ref{SecFSS}. 
\begin{itemize} 
 \item \textit{Method 1.} From Eq.\ (\ref{Taylor-espansion}), we computed $\bar{\varepsilon}^{(2)}_{+}$ 
 and $\bar{\varepsilon}^{(2)}_{-}$ at  $T^{(2)}_{+}=2.20177$ and $T^{(2)}_{-}= 2.201576$, respectively, 
 assuming $\varepsilon^{(2)}_{c}=-0.98904$ as reported in Eq.\ (\ref{XY_energy_I_best_result}). 
 These quantities are given by $\bar{\varepsilon}^{(2)}_{+}=-0.98629$ and $\bar{\varepsilon}^{(2)}_{-}=-0.99180$ 
 and are such that 
 $|\varepsilon^{(2)}_{c} - \bar{\varepsilon}^{(2)}_{+}| = |\varepsilon^{(2)}_{c}-\bar{\varepsilon}^{(2)}_{-}|\simeq 0.003$. 
 In this way, we get
\begin{equation}\label{XY_energy_II_best_result}
 \varDelta \bar{\varepsilon}^{(2),syst}_{c}= |\varepsilon^{(2)}_{c} - \bar{\varepsilon}^{(2)}_{\pm}|=0.003.
\end{equation}
 \item \textit{Method 2.} We computed $\tilde{\varepsilon}^{(2)}_{\pm}$ with a fit of the energy density data 
 $\tilde{\varepsilon}^{(2)}_{\pm}(L)$ for $L=40,50,80$ and $100$ 
 at $T^{(2)}_{+}=2.20177$ and $T^{(2)}_{-}= 2.201576$, respectively, according to Eq.\ (\ref{energy_FSS_out}) 
 with $n=2$ and $D_2=-1.5317$ as derived from data in \cite{GottlobHasenbusch:physicaa1993}. 
 $\tilde{\varepsilon}^{(2)}_{\pm}(L)$ for these values of $L$ are computed with Eq.\ (\ref{Taylor-espansion-2}) 
 from data given in Table \ref{table:xy_data_ec}.
 For some particular values of $L$, namely for $L=32, 64$ and $128$, we performed 
 Monte Carlo simulations at $T^{(2)}_{+}$ and $T^{(2)}_{-}$, respectively, 
 to compute the numerical values $\mathbf{\varepsilon}^{(2)}_{\pm}(32)$, $\mathbf{\varepsilon}^{(2)}_{\pm}(64)$ and 
 $\mathbf{\varepsilon}^{(2)}_{\pm}(128)$. The numerical results have been compared with the same quantities 
 as derived with the Taylor expansion (\ref{Taylor-espansion-2}) and appeared to be consistent with them. 
 This result reinforce the robustness of the analytical procedure used to derive 
 $\varDelta \tilde{\varepsilon}^{(2),syst}_{c}$ and we 
 considered the simulation values $\mathbf{\varepsilon}^{(2)}_{\pm}(32)$, $\mathbf{\varepsilon}^{(2)}_{\pm}(64)$ and 
 $\mathbf{\varepsilon}^{(2)}_{\pm}(128)$ in the fitting procedure for the derivation of 
 $\tilde{\varepsilon}^{(2)}_{\pm}$. 
 The data used in the analysis are given in Table \ref{table:xy_data_tc_piu_meno_delta} in which data derived 
 from Monte Carlo simulations are in \textbf{bold} and data derived with the Taylor expansion (\ref{Taylor-espansion-2}) are in plain text. The result of the fits are reported in Table \ref{xy_data_fit_delta_t}; we get 
\begin{equation}\label{XY_energy_III_best_result}
\varDelta \tilde{\varepsilon}^{(2),syst}_{c}= 
\cdot^{|\varepsilon^{(2)}_{c}-\tilde{\varepsilon}^{(2)}_{+}|}_{|\varepsilon^{(2)}_{c}-\varepsilon^{(2)}_{-}|}=
\cdot^{+0.0003}_{-0.0003}=0.0003\, .
\end{equation}
\end{itemize}

\begin{table}[ht]
\caption{ {Energy density data $\varepsilon^{(2)}_{+}$ and $\varepsilon^{(2)}_{-}$ obtained via Taylor expansion 
and numerical Monte Carlo simulations (\textbf{bold}), at $T^{(2)}_{+}=2.20177$ and $T^{(2)}_{-}=2.201576$, respectively.}}
\centering
\begin{tabular}{|c |c |c|}
\hline
$L$ & $\varepsilon^{(2)}_{+}(L)$ & $\varepsilon^{(2)}_{-}(L)$\\
[0.5ex]
\hline
\hline 
32 & \textbf{-0.99854(15)} & \textbf{-0.9984(3)} \\

40 & -0.99563(12) & -0.99615(12) \\

50 & -0.99355(9) & -0.99409(9)\\

64 &  \textbf{-0.99197(7)} & \textbf{-0.99270(7)} \\

80 & -0.99107(6) & -0.99167(6)\\ 

100 & -0.99036(4) & -0.99098(4)\\

128 &  \textbf{-0.98994(4)} & \textbf{-0.99049(4)} \\ [1ex] 
\hline
\end{tabular}
\label{table:xy_data_tc_piu_meno_delta}
\end{table}
\begin{table}[ht]
\caption{ {Fitting values of the parameters $\varepsilon^{(2)}_{\pm}$ and $\varepsilon^{(2)}_{\pm}$. In parentheses 
are the statistical errors due to the fitting procedure.}}
\centering
\begin{tabular}{|c| c| c| c|}
\hline
Fitting parameters & constants & results & $\chi^2/\text{d.o.f.}$ \\
[0.5ex]
\hline
\hline
 &	&  {$\varepsilon^{(2)}_{+}=-0.98871(5)$} &\\[-1ex]
\raisebox{1.5ex}{ {$\varepsilon^{(2)}_{+},\;\varepsilon_{2,+}$}}       & 
\raisebox{1.5ex}{ {$D_{2}=-1.5317$}}  &  {$\varepsilon_{2,+}=-1.95(3)$}       &
\raisebox{1.5ex}{ {$1.46$}} \\
\hline 

 &	&  {$\varepsilon^{(2)}_{-}=-0.98935(4)$} &\\[-1ex]
\raisebox{1.5ex}{ {$\varepsilon^{(2)}_{-},\;\varepsilon_{2,-}$}}       & 
\raisebox{1.5ex}{ {$D_2=-1.5317$}}  &  {$\varepsilon_{2,-}=-1.91(3)$}       &
\raisebox{1.5ex}{ {$0.8$}} \\
\hline 
\end{tabular}
\label{xy_data_fit_delta_t}
\end{table}

In Sec.\ \ref{Sec_NumericalTest} we are going 
to compare the critical values of the energy density of different O$(n)$ models 
both in the limit of small $n$ and in the limit $n\rightarrow\infty$; we should then consider $\varDelta \varepsilon^{(2),syst}_{c}=\varDelta \bar{\varepsilon}^{(2),syst}_{c} $ given 
in Eq.\ (\ref{XY_energy_II_best_result}), being the largest among the two different estimations of the systematic 
uncertainties reported in Eqs.\ (\ref{XY_energy_II_best_result}) and (\ref{XY_energy_III_best_result}), respectively.
However, this result depends on the value of $c^{(2)}_{c}$ given in Eq.\ (\ref{XY_cv_best_result}) that, in turn, 
is strongly affected by the choice of the ratio $\alpha_2/\nu_2$. 
For this reason we prefer to consider $\varDelta \tilde{\varepsilon}^{(2),syst}_{c}$ given in Eq.\ (\ref{XY_energy_III_best_result}) 
as best estimate of $\varDelta \varepsilon^{(2),syst}_{c}$. We finally have 
\begin{equation}\label{XY_energy_best_result}
\varepsilon^{(2)}_{c}\pm \varDelta \varepsilon^{(2),stat}_{c}\pm \varDelta \varepsilon^{(2),syst}_{c}=
-0.98904\;\pm\;0.00003\;\pm\;0.0003
\end{equation}
as final best estimate for the critical energy 
density of the O$(2)$ model in three dimensions. The uncertainty 
$\varDelta \varepsilon^{(2),syst}_{c}$ due to $\varDelta T^{(2)}_{c}$ 
is an order of magnitude larger than the statistical error: 
this feature will be in common with all the other 
models considered.

\subsection{$n=3$, the Heisenberg model}\label{numericalO3model}
We performed canonical Monte Carlo simulations of the Heisenberg model defined on a regular cubic lattices with 
edges $L=32,40,50,64,80,100$ and $128$. As best estimate of the critical temperature of the system 
we considered the value  $T^{(3)}_{c}=1.44298(2)$ given in \cite{BrownCiftan:prb2006}. 
The values for $\varepsilon^{(3)}_{c}(L)$ and $c^{(3)}_{c}(L)$ obtained from the simulations are reported in Table 
\ref{table:O3_data_ec}: in parentheses are the statistical errors.

\begin{table}[ht]
\caption{ {Monte Carlo results for the energy density $\varepsilon^{(3)}_{c})$ and for the specific heat 
$c^{(2)}_{c}$ at the critical temperature $T^{(3)}_{c}=1.44298$.}}
\centering
\begin{tabular}{|c |c |c|}
\hline
$L$ & $\varepsilon^{(3)}_{c}(L)$ & $c^{(3)}_{c}(L)$ \\
[0.5ex]
\hline
\hline 
32 & -0.99646(7) & 2.863(15) \\ 

40 & -0.99437(6) & 2.938(19) \\


50 & -0.99289(5) & 3.030(19) \\


64 & -0.99183(4) & 3.126(23) \\


80 & -0.99116(3) & 3.197(28) \\

100 & -0.99064(3) & 3.259(32) \\

128 & -0.990312(14) & 3.367(28) \\ [1ex] 
\hline
\end{tabular}
\label{table:O3_data_ec}
\end{table}

We fitted data reported in Table \ref{table:O3_data_ec} according to 
relation (\ref{energy_FSS}) with $n=3$ and 
considering $\varepsilon^{(3)}_{c}$ and $\varepsilon_{3}$ as fitting parameters.
For the values of the critical exponents, we considered different choices: 
(\textit{i}) the best theoretical estimates $\nu_3=0.705(3)$ and $\alpha_3=-0.115(9)$ coming from a 
re-summed perturbation series analysis \cite{LeGuillouZinnJustin:prb1980}; 
(\textit{ii}) we used $\alpha_{3}-1)/\nu_{3}=-1.586(19)$ as obtained in \cite{HolmJanke:jphysa1994} from a similar analysis performed using a 
slightly different value of the critical temperature, namely $T_{c}=1.4430$; 
(\textit{iii}) we considered $(\alpha_{3}-1)/\nu_{3}=-1.5974$ as derived 
in \cite{BrownCiftan:prb2006} from a similar analysis performed using t
he same value of $T^{(3)}_{c}$ as in our case. 
The results of these fits for $\varepsilon^{(3)}_{c}$ and $\varepsilon_{3}$ are reported in Table \ref{table:O3_fit_results}. 

\begin{table}[ht]
\caption{ {Fitting values of the parameters $\varepsilon^{(3)}_{c}$ and $\varepsilon_{3}$ 
entering expression (\ref{energy_FSS}).}}
\centering
\begin{tabular}{|c| c| c| c|}
\hline
Fitting parameters & $\nu_3$ and $\alpha_3$ & results & $\chi^2/\text{d.o.f.}$ \\
[0.5ex]
\hline
\hline
 &  {$\nu_3=0.705$}     &  {$\varepsilon^{(3)}_{c}= -0.989537(12)$} &\\[-1ex]
\raisebox{1.5ex}{ {$\varepsilon^{(3)}_{c},\;\varepsilon_{3}$}}       & 
 {$\alpha_3=-0.115$} &  {$\varepsilon_{3}=-1.652(10)$}       &
\raisebox{1.5ex}{ {$0.52$}} \\
\hline 

 &      &  {$\varepsilon^{(3)}_{c}=-0.989542(11)$} & \\[-1ex]
\raisebox{1.5ex}{ {$\varepsilon^{(3)}_{c},\;\varepsilon_{3}$}}  & 
\raisebox{1.5ex}{ {$D_{3}=-1.586$}} &  {$\varepsilon_{3}=-1.677(10)$}       &
\raisebox{1.5ex}{ {$0.48$}} \\
\hline 

 &      &  {$\varepsilon^{(3)}_{c}=-0.989556(10)$} & \\[-1ex]
\raisebox{1.5ex}{ {$\varepsilon^{(3)}_{c},\;\varepsilon_{3}$}}  & 
\raisebox{1.5ex}{ {$D_{3}=-1.5974$}} &  {$\varepsilon_{3}=-1.744(9)$}       &
\raisebox{1.5ex}{ {$0.40$}} \\
\hline 
\end{tabular}
\label{table:O3_fit_results}
\end{table}

We also performed a fit of all the parameters 
$\varepsilon^{(3)}_{c}$, $\varepsilon_{3}$ and $D_{3} = (\alpha_3 - 1)/\nu_3$ 
with the scaling relation $\varepsilon^{(3)}_{c}(L)=\varepsilon^{(3)}_{c}+\varepsilon_{3}L^{D_{3}}$. 
The results are $\varepsilon^{(3)}_{c}=-0.98958(3)$, $\varepsilon_{3}=-1.88(17)$ and $D_{3}=-1.62(2)$ with a 
$\chi^2/\text{d.o.f.}\simeq 0.43$. 
These results are in agreement with those reported in Table \ref{table:O3_fit_results} and with 
the results reported in literature, see e.g.\ \cite{HolmJanke:jphysa1994,BrownCiftan:prb2006}. However, 
as they come from a three-parameters fit of a relatively small set of data, 
we chose to neglect them and 
to consider only results reported in Table \ref{table:O3_fit_results} 
in our study. 

The values of the parameters reported in the second 
row of Table \ref{table:O3_fit_results} 
are consistent with the corresponding quantities 
reported in \cite{HolmJanke:jphysa1994}. 
Therein, the authors obtain  
$\varepsilon^{(3)}_{c}=-0.9894(1)$, $\varepsilon_{3}=-1.68(8)$ and $D_{3}=-1.586(19)$. 
These values come from a three parameter fit of the scaling relation 
$\varepsilon^{(3)}_{c}(L) = \varepsilon^{(3)}_{c} + \varepsilon_{3} L^{D_{3}}$ with $D_{3}=(\alpha_3-1)/\nu_3$, 
performed at $T_{c}=1.4430\neq T^{(3)}_{c}$. Beside supporting our results, 
this fact seems to suggest that $\varepsilon^{(3)}_{c}$ does not sensibly depend 
on the value of the 
critical temperature. 

For what concerns the third row of Table \ref{table:O3_fit_results}, 
the results of the fit 
have to be compared with the results computed 
in \cite{BrownCiftan:prb2006} at the same value 
of $T^{(3)}_{c}$ as in our case. 
Therein, the authors find
\begin{equation}
\varepsilon^{(3)}_{c}(L)=\varepsilon^{(3)}_{c}+\varepsilon_{3} L^{D_{3}} = -0.9896\pm1.7225 L^{-1.5974}  \, ; 
\end{equation}
the relative precision of the data fit being of $0.001\%$ or better. 
Also in this case our results, obtained for $D_{3}=-1.5974$, are perfectly 
consistent.

\begin{figure}
\centering
\includegraphics[width=10cm]{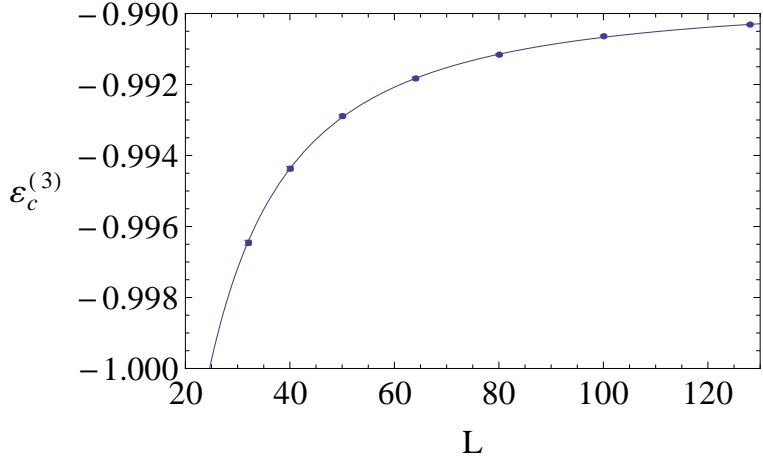} 
\caption{ {Energy density $\varepsilon^{(3)}_{c}$ at the critical temperature $T^{(3)}_{c}=1.4498$ as a 
function of $L$. The solid curve is the fit to (\ref{H_energy_I_best_result}) with $(\alpha_3-1)/\nu_3=-1.5974$.}}
\label{H_energy_plot}
\end{figure}

The values of the parameter $\varepsilon^{(3)}_{c}$ reported in Table \ref{table:O3_fit_results} are 
consistent with each other. The results reported in the third row of Table \ref{table:O3_fit_results} have been determined 
considering a combination of the critical exponents $D_{3}$ as derived in \cite{BrownCiftan:prb2006} at the same value of 
the critical temperature as in our case. 
Since the numerical value of $\alpha_3/\nu_{3}$ is needed in the following 
to determine $c^{(3)}_{c}$, we give
\begin{equation}\label{H_energy_I_best_result}
\begin{split}
\varepsilon^{(3)}_{c}\;\pm\;\varDelta \varepsilon^{(3),stat}_{c} &=-0.989556\;\pm\;0.000010\,,\\ 
\varepsilon_{3}&=-1.744\;\pm\;0.009\, ;
\end{split}
 \end{equation}
as best estimate of the critical energy density value of $\varepsilon^{(3)}_{c}$. 
The curve $\varepsilon^{(3)}_{c}(L)$ given 
by Eq.\ (\ref{energy_FSS}) for $n=3$ and with the values of 
$\varepsilon^{(3)}_{c}$ and $\varepsilon_{3}$ as in Eq.\ (\ref{H_energy_I_best_result}) is shown in Fig. \ref{H_energy_plot} 
together with the simulation data. 
It is worth noticing that the value of $\varepsilon^{(3)}_{c}$ 
in Eq.\ (\ref{H_energy_I_best_result}) 
is given with one digit of precision more than previous results 
in the literature and obtained with similar techniques 
\cite{HolmJanke:jphysa1994,BrownCiftan:prb2006}.

\begin{figure}
\centering
\includegraphics[width=10cm]{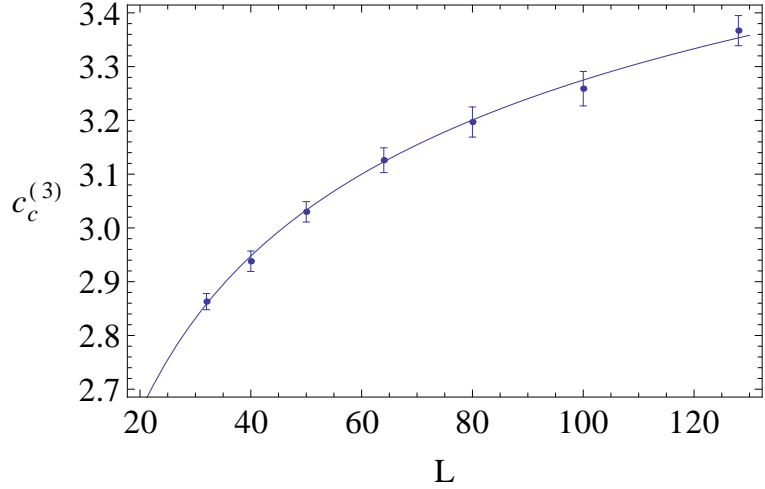} 
\caption{ {Specific heat $c^{(3)}_{c}$ at the critical temperature $T^{(3)}_{c}=1.4498$ as a 
function of $L$. The solid curve is the fit to (\ref{H_cv_best_results}) with $\alpha_3/\nu_3=-0.1991$. }}
\label{H_cv_plot}
\end{figure}

We fitted data of $c^{(3)}_{c}(L)$ reported in Table \ref{table:O3_data_ec} according to the scaling relation given 
in Eq.\ (\ref{cv_FSS}) with $\alpha_3/\nu_3=-0.1991$ as in \cite{BrownCiftan:prb2006}. The results of the fit 
are reported in the first row of Table \ref{table:O3_cv_fit_results}. 
To check the dependence of our results from the ratio $\alpha_3/\nu_3$ we performed the same fit for 
two different choices of $\alpha_3/\nu_3$: 
(\textit{i}) $\alpha_3/\nu_3=-0.1631$ as derived in \cite{LeGuillouZinnJustin:prb1980} and 
(\textit{ii}) $\alpha_3/\nu_3=-0.166$ as derived in \cite{HolmJanke:jphysa1994}. 
The results of these fits are reported in the second and third rows of Table \ref{table:O3_cv_fit_results}, respectively. 
We chose 
\begin{equation}\label{H_cv_best_results}
\begin{split}
c^{(3)}_{c}&=4.91\;\pm\;0.03\,,\\
c_{3} &=-4.09\;\pm\;0.09\,;
\end{split}
\end{equation}
as the best choice of the fitting parameters, being associated to a choice of the critical exponents as in 
\cite{BrownCiftan:prb2006} at the same value of $T^{(3)}_{3}$ as in our case. 
The curve $c^{(3)}_{c}(L)$ given 
by Eq.\ (\ref{cv_FSS}) for $n=3$ 
and the values of the fitting parameters $c^{(3)}_{c}$ and $c_{3}$ as in 
Eq.\ (\ref{H_cv_best_results}) is shown in Fig.\ \ref{H_cv_plot} 
together with the simulation data. 

\begin{table}[ht]
\caption{ {Fitting values of the parameters $c^{(3)}_{c}$ and $c_{3}$ entering expression (\ref{energy_FSS}) 
with $n=3$.}}
\centering
\begin{tabular}{|c| c| c| c|}
\hline
Fitting parameters & constants & results & $\chi^2/\text{d.o.f.}$ \\
[0.5ex]
\hline
\hline
 &	&  {$c^{(3)}_{c}=4.91(3)$} &\\[-1ex]
\raisebox{1.5ex}{ {$c^{(3)}_{c},\;c_{3}$}}       & 
\raisebox{1.5ex}{ {$\alpha_3/\nu_3=-0.1991$}}  &  {$c_{2}=-4.09(9)$}       &
\raisebox{1.5ex}{ {$0.18$}} \\
\hline 

 & 	&  {$c^{(3)}_{c}=5.31(5)$} &\\[-1ex]
\raisebox{1.5ex}{ {$c^{(3)}_{c},\;c_{3}$}}       & 
\raisebox{1.5ex}{ {$\alpha_3/\nu_3=-0.1631$}} &  {$c_{3}=-4.32(8)$}       &
\raisebox{1.5ex}{ {$0.15$}} \\
\hline 

 & 	&  {$c^{(3)}_{c}=5.27(4)$} &\\[-1ex]
\raisebox{1.5ex}{ {$c^{(3)}_{c},\;c_{3}$}}       & 
\raisebox{1.5ex}{ {$\alpha_3/\nu_3=-0.166$}} &  {$c_{3}=-4.29(8)$}       &
\raisebox{1.5ex}{ {$0.15$}} \\
\hline 
\end{tabular}
\label{table:O3_cv_fit_results}
\end{table}

In order to evaluate $\varDelta \varepsilon^{(3),syst}_{c}$, 
we applied the two methods presented in 
Sec. \ref{SecFSS} specialized to $n=3$:
\begin{itemize}
 \item \textit{Method 1.} From Eq. (\ref{Taylor-espansion}) we computed the values of $\bar{\varepsilon}^{(3)}_{+}$ 
 and $\bar{\varepsilon}^{(3)}_{-}$ at $T^{(3)}_{+}=1.44300$ and $T^{(3)}_{-}= 1.44296$, respectively, 
 assuming $\varepsilon^{(3)}_{c}=-0.989556$ as reported in Eq.\  (\ref{H_energy_I_best_result}). 
 These quantities are given by $\bar{\varepsilon}^{(3)}_{+}=-0.989458$ and $\bar{\varepsilon}^{(3)}_{-}=-0.989654$
 and are such that 
 $|\varepsilon^{(3)}_{c} - \bar{\varepsilon}^{(3)}_{+}| = |\varepsilon^{(3)}_{c}-\bar{\varepsilon}^{(3)}_{-}|\simeq 0.00010$. 
 In this way, we get
\begin{equation}\label{H_energy_II_best_result}
 \varDelta \bar{\varepsilon}^{(3),syst}_{c}= |\varepsilon^{(3)}_{c} - \bar{\varepsilon}^{(3)}_{\pm}|=0.00010.
\end{equation}
\item \textit{Method 2.} We computed $\tilde{\varepsilon}^{(3)}_{\pm}$ with a fit of the energy density data 
 for $\tilde{\varepsilon}^{(3)}_{\pm}(L)$ for $L=32,40,50,64,80,100$ and $128$ 
 at $T^{(3)}_{+}=1.44300$ and $T^{(3)}_{-}= 1.44296$, respectively, according to Eq.\ (\ref{energy_FSS_out}) 
 with $n=3$ and $D_{3}=-1.5974$ as in \cite{BrownCiftan:prb2006}. For $L=40,50,80,100$ we computed 
 $\tilde{\varepsilon}^{(3)}_{\pm}(L)$ by applying Eq.\ (\ref{Taylor-espansion-2}) to data 
 given in Table \ref{table:O3_data_ec}. As in the case of the $XY$ model, the values of
 $\tilde{\varepsilon}^{(3)}_{\pm}(L)$ for $L=32,64$ and $128$ are obtained with Monte Carlo simulations performed at 
 $T^{(3)}_{+}$ and  $T^{(3)}_{-}$, respectively; these numerical values are consistent with the same quantities computed with 
 Eq.\ (\ref{Taylor-espansion-2}), not shown here. The data involved in the analysis are shown in 
 Table \ref{table:H_data_tc_piu_meno_delta}; 
 data arising from the Monte Carlo simulations are printed in \textbf{bold} and data computed using Eq.\ (\ref{Taylor-espansion-2}) are printed in plain text.
 From the fits we get
\begin{equation}\label{H_energy_III_best_result}
 \varDelta \tilde{\varepsilon}^{(3),syst}_{c}= 
\cdot^{|\varepsilon^{(3)}_{+} - \varepsilon^{(3)}_{c}|}_{|\varepsilon^{(3)}_{-}-\varepsilon^{(3)}_{c}|}=
\cdot^{+0.00008}_{-0.00006}
\end{equation}
\end{itemize}
as reported in Table \ref{O3_data_fit_delta_t}. 
Since our purpose is to compare the values of the critical energy density for different O$(n)$ models, we choose 
$\varDelta \bar{\varepsilon}^{(3),syst}_{c}$ in Eq.\ (\ref{H_energy_II_best_result}) as best estimate of the 
systematic uncertainty on $\varepsilon^{(3)}_{c}$. From Eqs.\ (\ref{H_energy_I_best_result}) and 
(\ref{H_energy_II_best_result}) 
we finally get 
\begin{equation}\label{H_energy_final_best_result}
  \varepsilon^{(3)}_{c}\;\pm\;\varDelta \varepsilon^{(3),stat}_{c}\;\pm\;\varDelta \varepsilon^{(3),syst}_{c}=
  -0.989556\;\pm\;0.000010\;\pm\;0.00010,
\end{equation}
as best estimate of the critical energy density of the three dimensional Heisenberg model, in the thermodynamic limit.
\begin{table}[ht]
\caption{ {Energy density data $\varepsilon^{(3)}_{+}(L)$ and $\varepsilon^{(3)}_{-}(L)$ obtained via 
Taylor expansion (plain text) and numerical Monte Carlo simulations (\textbf{bold}), at 
$T^{(3)}_{+}=1.44300$ and $T^{(3)}_{-}=1.44296$, respectively. The statistical errors are in parentheses.}}
\centering
\begin{tabular}{|c |c |c|}
\hline
$L$ & $\varepsilon^{(3)}_{+}(L)$ & $\varepsilon^{(3)}_{-}(L)$\\[0.5ex]
\hline
\hline 
32 & \textbf{-0.99636(7)} & \textbf{-0.99654(7)}\\
40 & -0.99431 & -0.99443\\
50 &-0.99283 & -0.99295\\
64 & \textbf{-0.99164(6)}& \textbf{-0.99182(4)}\\
80 & -0.99110& -0.99122\\
100 & -0.99058& -0.99071\\
128 & \textbf{-0.990232(19)}& \textbf{-0.99039(2)} \\ [1ex] 
\hline
\end{tabular}
\label{table:H_data_tc_piu_meno_delta}
\end{table}
\begin{table}[ht]
\caption{ {Fitting values of the parameters $\varepsilon^{3}_{\pm}$ and $\varepsilon_{\pm,3}$.}}
\centering
\begin{tabular}{|c| c| c| c|}
\hline
Fitting parameters & $D_3$ & results & $\chi^2/\text{d.o.f.}$ \\
[0.5ex]
\hline
\hline
 &	&  {$\varepsilon^{(3)}_{+}=-0.989479(19)$} &\\[-1ex]
\raisebox{1.5ex}{ {$\varepsilon^{(3)}_{+},\;\varepsilon_{+,3}$}}       & 
\raisebox{1.5ex}{ {$D_3=-1.5974$}}  &  {$\varepsilon_{+,3}=-1.743(16)$}       &
\raisebox{1.5ex}{ {$0.97$}} \\
\hline 

 &	&  {$\varepsilon^{(3)}_{-}=-0.98962(2)$} &\\[-1ex]
\raisebox{1.5ex}{ {$\varepsilon^{(3)}_{-},\;\varepsilon_{-,3}$}}       & 
\raisebox{1.5ex}{ {$D_3=-1.5974$}}  &  {$\varepsilon_{-,3}=-1.738(17)$}       &
\raisebox{1.5ex}{ {$1.15$}} \\
\hline 
\end{tabular}
\label{O3_data_fit_delta_t}
\end{table}

\subsection{$n=4$, the O$(4)$ model}\label{numericalO4model}
We performed canonical Monte Carlo simulations of the O$(4)$ model on a regular cubic lattices with 
edges $L=32,40,64,80,100$ and $128$. For the critical temperature of the system we choose the value  
$T^{(4)}_{c}=1.06835(13)$ given in \cite{KanayaKaya:prd1995}; 
therefore, simulations were performed at $T=1.06835$. 
Table \ref{table:O4_data_ec} shows the values for $\varepsilon^{(4)}_{c}(L)$ and $c^{(4)}_{c}(L)$ involved in the analysis, with statistical errors in parentheses. 

\begin{table}[ht]
\caption{ {Monte Carlo results for the energy density $\varepsilon^{(4)}_{c}(L)$ and for the specific heat 
$c^{(4)}_{c}(L)$ at the critical temperature $T^{(4)}_{c}=1.06835$.}}
\centering
\begin{tabular}{|c |c |c|}
\hline
$L$ & $\varepsilon^{(4)}_{c}(L)$ & $c^{(4)}_{c}(L)$ \\
[0.5ex]
\hline
\hline 
32 & -0.996930(67) & 3.195(20) \\ 

40 & -0.995431(53) & 3.282(21) \\

64 & -0.993374(35) & 3.416(27) \\

80 & -0.992875(20) & 3.470(39) \\

100 & -0.992482(23) & 3.551(44)\\

128 & -0.992260(20) & 3.617(43)\\ [1ex] 
\hline
\end{tabular}
\label{table:O4_data_ec}
\end{table}

We fitted data reported in Table \ref{table:O4_data_ec} according to Eq.\ (\ref{energy_FSS}) with $n=4$ 
and considering $\varepsilon^{(4)}_{c}$ and $\varepsilon_{4}$ as fitting parameters. 
For the values of the critical exponents, we considered two different cases: 
(\textit{i}) $\nu_4=0.7479(80)$ as reported in \cite{KanayaKaya:prd1995} using the same value of the critical temperature as 
in our case and $\alpha_4=-0.244(24)$ as obtained from the scaling relation $\alpha=2-d\nu$ with $d=3$; 
(\textit{ii}) $\alpha_4=-0.21312$ and $\nu_4=0.73771$ as obtained from the scaling relations 
$\alpha=2 - \beta ( 1 + \delta )$ and $\nu=\frac{2-\alpha}{d}$ with $d=3$, from data reported in \cite{EngelsKarsch:prd2012} using 
$T_{c}=1.06849$. In \cite{EngelsKarsch:prd2012} the values of 
$\varepsilon^{(4)}_{c}$ and $c^{(4)}_{c}$ have been determined with a finite size scaling analysis in an external 
field $h$ and then extrapolating the results in the limit $h\rightarrow 0$. As we shall see in the following, their results are in good agreement with ours although derived with a slightly different approach: this supports the validity of  
our analysis.
The results of the fits for $\varepsilon^{(4)}_{c}$ and $\varepsilon_{4}$ are reported in Table \ref{table:O4_fit_results}.

\begin{table}[ht]
\caption{ {Fitting values of the parameters $\varepsilon^{(4)}_{c}$ and $\varepsilon_{4}$ 
entering Eq.\ (\ref{energy_FSS}).}}
\centering
\begin{tabular}{|c| c| c| c|}
\hline
Fitting parameters & $\nu_4$ and $\alpha_4$ & results & $\chi^2/\text{d.o.f.}$ \\
[0.5ex]
\hline
\hline
 &  {$\nu_4=0.7479$}     &  {$\varepsilon^{(4)}_{c}= -0.99174(2)$} &\\[-1ex]
\raisebox{1.5ex}{ {$\varepsilon^{(4)}_{c},\;\varepsilon_{4}$}}       & 
 {$\alpha_4=-0.244$} &  {$\varepsilon_{4}=-1.68(2)$}       &
\raisebox{1.5ex}{ {$1.3$}} \\
\hline 

 &  {$\nu_4=0.73771$}     &  {$\varepsilon^{(4)}_{c}= -0.99170(2)$} & \\[-1ex]
\raisebox{1.5ex}{ {$\varepsilon^{(4)}_{c},\;\varepsilon_{4}$}}  &{ {$\alpha_4=-0.21312$}} & 
 {$\varepsilon_{4}=-1.57(2)$}       &\raisebox{1.5ex}{ {$1.3$}} \\
\hline 
\end{tabular}
\label{table:O4_fit_results}
\end{table}
We also performed a four-parameter fit with $\alpha_4$, $\nu_4$, $\varepsilon^{(4)}_{c}$ and $\varepsilon_{4}$ as free parameters. 
However, as in the $n=2$ case, no meaningful results can be extracted from the fit, the relative error on the 
critical exponents being larger then $100\%$. The results of the fit are not shown here and will be neglected in 
the following. 

\begin{figure}
\centering
\includegraphics[width=10cm]{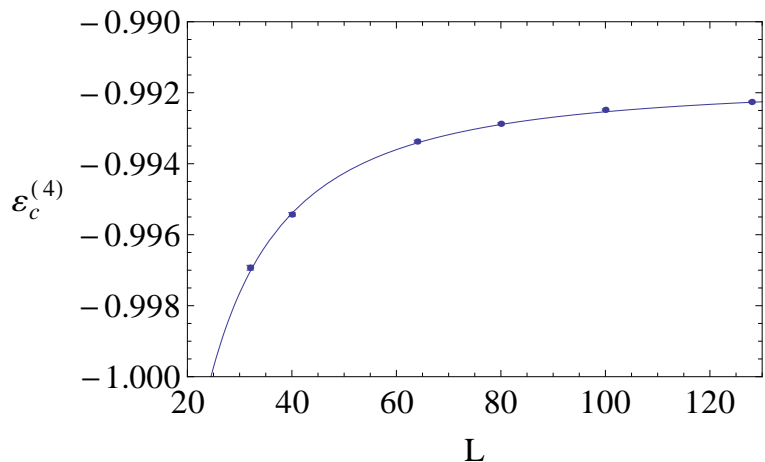} 
\caption{ {Energy density $\varepsilon^{(4)}_{c}$ at the critical temperature $T^{(4)}_{c}=1.06835$ as a function of $L$. The solid curve is the fit to Eq.\ (\ref{energy_FSS}) with $\alpha_4=-0.244$ and $\nu_4=0.7479$.}}\label{O4_energy_plot}
\end{figure}

The results for the critical energy density $\varepsilon^{(4)}_{c}$ shown in Table \ref{table:O4_fit_results} 
are consistent with each other. 
As anticipated, they are also in good agreement with the known results, see e.g.\ \cite{EngelsKarsch:prd2012}, where the 
authors find $\varepsilon^{(4)}_{c}=-0.991792(28)$ from a FSS analysis in an external magnetic field. 
We chose 
\begin{equation}\label{O4_energy_I_best_result}
\begin{split}
\varepsilon^{(4)}_{c}\;\pm\;\varDelta \varepsilon^{(4),stat}_{c}&= -0.99174\;\pm\;0.00002\,,\\ 
\varepsilon_{4}&=-1.69\;\pm\;0.02
\end{split}
\end{equation}
as best estimate of the critical energy density $\varepsilon^{(4)}_{c}$ and of the fitting parameter 
$\varepsilon_{4}$, as reported in the first row of Table \ref{table:O4_fit_results}. Indeed, these results come from a 
choice of the critical exponents as in \cite{KanayaKaya:prd1995} where the same value of the critical temperature as 
in our case was used. The curve $\varepsilon^{(4)}_{c}(L)$ given by Eq.\ (\ref{energy_FSS}) for $n=3$ and for $\varepsilon^{(4)}_{c}$ and $\varepsilon_{4}$ as in Eq.\ (\ref{O4_energy_I_best_result}), 
is shown in Fig.\ \ref{O4_energy_plot} together with the simulation data used in the analysis.

\begin{figure}
\centering
\includegraphics[width=10cm]{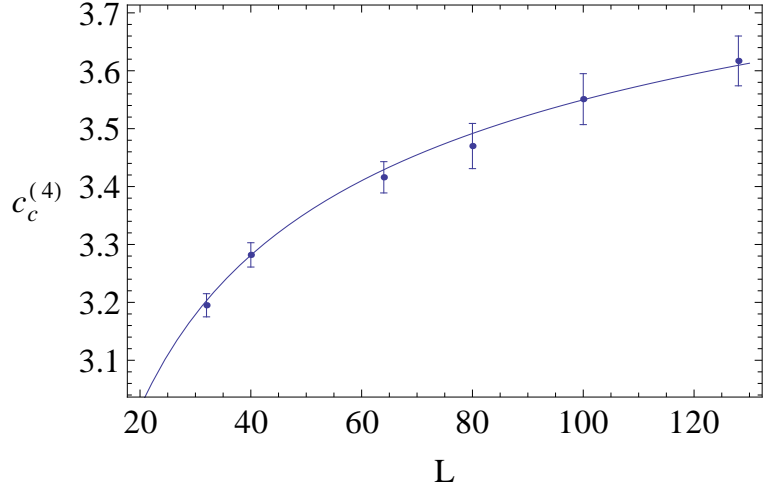} 
\caption{ {Specific heat $c^{(4)}_{c}$ at the critical temperature  $T^{(4)}_{c}=1.06835$ as a function of $L$. 
The solid curve is the fit to (\ref{cv_FSS}) 
with $\alpha_4/\nu_4=-0.326$.}}\label{O4_cv_plot}
\end{figure}

We fitted data of $c^{(4)}_{c}(L)$ reported in Table \ref{table:O4_data_ec} according to the scaling relation given
in Eq.\ (\ref{cv_FSS}) with $n=4$ and keeping the value of the ratio $\alpha_4/\nu_4$ fixed to $\alpha_4/\nu_4=-0.326$ 
as derived in \cite{KanayaKaya:prd1995} at the same value of $T^{(4)}_{c}$ as in our case. 
The results of the fit are given by 
\begin{equation}\label{O4_cv_best_results}
\begin{split}
c^{(4)}_{c}&=4.32\;\pm\;0.03\,,\\
c_{4}&=-3.46\;\pm\;0.10\,,
\end{split}
\end{equation}
and are reported in the first row of Table \ref{table:O4_cv_fit_results}. 
To check the dependence of our results on the value of the ratio $\alpha_4/\nu_4$, 
we also performed the fit with a different choice for $\alpha_4/\nu_4$: 
$\alpha_4/\nu_4=-0.289$ as derived from data reported in \cite{EngelsKarsch:prd2012}. 
The results of this fit are reported in the second row of Table \ref{table:O4_cv_fit_results}. 
The values of $c^{(4)}_{c}$ reported in Table \ref{table:O4_cv_fit_results} are in a good agreement with each other. 
Moreover the value of $c^{(4)}_{c}$ in the second row Table \ref{table:O4_cv_fit_results} is 
consistent with the corresponding quantity reported in \cite{EngelsKarsch:prd2012} and derived with a rather different procedure.
\\
\begin{table}[ht]
\caption{ {Fitting values of the parameters $c^{(4)}_{c}$ and $c_{4}$ entering expression (\ref{energy_FSS}) with 
$n=4$.}}
\centering
\begin{tabular}{|c| c| c| c|}
\hline
Fitting parameters & constants & results & $\chi^2/d.o.f$ \\
[0.5ex]
\hline
\hline
 &	&  {$c^{(4)}_{c}=4.32(3)$} &\\[-1ex]
\raisebox{1.5ex}{ {$c^{(4)}_{c},\;c_{4}$}}       & 
\raisebox{1.5ex}{ {$\alpha_4/\nu_4=-0.326$}}  & 
 {$c_{4}=-3.46(10)$}       &\raisebox{1.5ex}{ {$0.12$}} \\
\hline 

 & 	&  {$c^{(4)}_{c}=4.43(3)$} &\\[-1ex]
\raisebox{1.5ex}{ {$c^{(4)}_{c},\;c_{4}$}}       & 
\raisebox{1.5ex}{ {$\alpha_4/\nu_4=-0.289$}} & 
 {$c_{4}=-3.37(9)$}       & \raisebox{1.5ex}{ {$0.11$}} \\
\hline 
\end{tabular}
\label{table:O4_cv_fit_results}
\end{table}

In order to estimate $\varDelta \varepsilon^{(4),syst}_{c}$ we applied the two methods presented in Sec.\ \ref{SecFSS}:
\begin{itemize}
 \item \textit{Method 1.} From Eq.\ (\ref{Taylor-espansion}), we computed the values of $\bar{\varepsilon}^{(4)}_{+}$ 
 and $\bar{\varepsilon}^{(4)}_{-}$ at $T^{(4)}_{+}=1.06848$ and $T^{(4)}_{-}=1.06822$, respectively, 
 assuming $\varepsilon^{(4)}_{c}=-0.99174$ as reported in Eq.\ (\ref{O4_energy_I_best_result}). 
 These quantities are given by $\bar{\varepsilon}^{(4)}_{+}=-0.991178$ and $\bar{\varepsilon}^{(4)}_{-}=-0.992302$
 and are such that 
 $|\varepsilon^{(4)}_{c} - \bar{\varepsilon}^{(4)}_{+}| = |\varepsilon^{(4)}_{c}-\bar{\varepsilon}^{(4)}_{-}|\simeq 0.0006$. 
 In this way, we get
\begin{equation}\label{O4_energy_II_best_result}
 \varDelta \bar{\varepsilon}^{(4),syst}_{c}= |\varepsilon^{(4)}_{c} - \bar{\varepsilon}^{(4)}_{\pm}|=0.0006.
\end{equation}
 \item \textit{Method 2.} We computed $\tilde{\varepsilon}^{(4)}_{\pm}$ with a fit of the energy density data 
 $\tilde{\varepsilon}^{(4)}_{\pm}(L)$ with $L=32,64$ and $128$ derived with Monte Carlo simulations 
 performed at $T^{(4)}_{+}=1.06848$ and $T^{(4)}_{-}=1.06822$, respectively; 
 the fits have been computed according to relation in Eq.\ (\ref{energy_FSS_out}) with $n=4$ and 
 $D_{4}=-0.326$ as in \cite{KanayaKaya:prd1995}. At variance with what we have done for $n=2$ and $3$, in this case we 
 did not consider the values of the critical energy density for other $L$-values, obtained 
 with Eq.\ (\ref{Taylor-espansion-2}). 
 Indeed, in this case, the fits produced extremely bad results when Taylor-expanded data are considered.
 The Monte Carlo data involved in the analysis are given in Table \ref{table:O4_data_tc_piu_meno_delta}; 
 the statistical errors are reported in parentheses. The results of the fit, shown in Table \ref{O4_data_fit_delta_t}, are 
 such that
 \begin{equation}\label{O4_energy_III_best_result}
 \varDelta \tilde{\varepsilon}^{(4),syst}_{c}= 
 \cdot^{|\varepsilon^{(4)}_{+} - \varepsilon^{(4)}_{c}|}_{|\varepsilon^{(4)}_{-}-\varepsilon^{(4)}_{c}|}=
 \cdot^{+0.00006}_{-0.00002}
 \end{equation}
\end{itemize}
As for the O$(2)$ and for the O$(3)$ model, we are going to consider 
$\varDelta \varepsilon^{(4),syst}_{c}=\varDelta \bar{\varepsilon}^{(4),syst}_{c}=0.0006$ given by 
Eq.\ (\ref{O4_energy_II_best_result}), being larger than $\varDelta \tilde{\varepsilon}^{(4),syst}_{c}$ 
reported in Eq.\ (\ref{O4_energy_III_best_result}).

We finally get 
\begin{equation}\label{O4_energy_final_best_result}
  \varepsilon^{(4)}_{c}\;\pm\;\varDelta \varepsilon^{(4),stat}_{c}\;\pm\;\varDelta \varepsilon^{(4),syst}_{c}=
  -0.99174\;\pm\;0.00002 \;\pm\;0.0006
\end{equation}
as the final value of the critical energy density of the three dimensional O$(4)$ model in the thermodynamic limit. As for the 
O$(2)$ and the O$(3)$ models, the uncertainty on $\varepsilon^{(4)}_{c}$ due to $\varDelta T^{(4)}_{c}$ is larger than the 
statistical uncertainty. 

\begin{table}[ht]
\caption{ {Energy density data $\varepsilon^{(4)}_{+}(L)$ and $\varepsilon^{(4)}_{-}(L)$ obtained with
numerical Monte Carlo simulations performed at $T^{(4)}_{+}=1.06848$ and $T^{(4)}_{-}=1.06822$, respectively.}}
\centering
\begin{tabular}{|c |c |c|}
\hline
$L$ & $\varepsilon^{(4)}_{+}(L)$ & $\varepsilon^{(4)}_{-}(L)$\\
[0.5ex]
\hline
\hline 
32 & -0.996955(64) & -0.996962(67)\\
64 & -0.993294(37)& -0.993383(36)\\
128 & -0.992208(19)& -0.992275(18)\\ [1ex] 
\hline
\end{tabular}
\label{table:O4_data_tc_piu_meno_delta}
\end{table}
\begin{table}[ht]
\caption{ {Fitting values of the parameters $\varepsilon^{(4)}_{\pm}$ and $\varepsilon_{4,\pm}$.}}
\centering
\begin{tabular}{|c| c| c| c|}
\hline
Fitting parameters & $D_4$ & results & $\chi^2/\text{d.o.f.}$ \\
[0.5ex]
\hline
\hline
 &	&  {$\varepsilon^{(4)}_{+}=-0.99168(3)$} &\\[-1ex]
\raisebox{1.5ex}{ {$\varepsilon^{(4)}_{+},\;\varepsilon_{4,+}$}}       & 
\raisebox{1.5ex}{ {$D_{4}=-0.326$}}  &  {$\varepsilon_{4,+}=-1.67(3)$}       &
\raisebox{1.5ex}{ {$1.5$}} \\
\hline 

 &	&  {$\varepsilon^{(4)}_{-}=-0.991755(8)$} &\\[-1ex]
\raisebox{1.5ex}{ {$\varepsilon^{(4)}_{-},\;\varepsilon_{4,-}$}}       & 
\raisebox{1.5ex}{ {$D_4=-0.326$}}  &  {$\varepsilon_{4,-}=-1.657(9))$}       &
\raisebox{1.5ex}{ {$0.16$}} \\
\hline 
\end{tabular}
\label{O4_data_fit_delta_t}
\end{table}

\subsection {$n=\infty$, the spherical model}\label{numericalSphericalModel}
The spherical model has been introduced by Berlin and Kac 
\cite{BerlinKac:physrev1952} as an exactly solvable model of a ferromagnet: 
its Hamiltonian reads
\begin{equation}\label{Ham_sph_model}
 H^{sph}= - \sum_{\langle i,j \rangle}^N T_i T_j\, ,
\end{equation}
where the sum is intended over all the distinct pairs of 
distinct nearest neighbors on a regular $d-$dimensional hypercubic lattice. 
At variance with the O$(n)$ models, 
the ``spin variables'' $T_i$ are real numbers 
and their modulus is not fixed to unity: instead, the spherical constraint 
\begin{equation}\label{sph_constraint}
\sum_{i=1}^{N} T^{2}_{i}=N
\end{equation}
is imposed, allowing for a fluctuation of the modulus of the spin variables. 

The spherical model is exactly solvable in any spatial dimension $d$ in 
the thermodynamic limit, both in the canonical and in the 
microcanonical ensembles: 
for the canonical solution see e.g.\ \cite{Binney:book} and references therein, 
for the microcanonical solution see \cite{Kastner:jstat2009}. 
Despite the long-range nature of the constraint 
in Eq.\ (\ref{sph_constraint}) the canonical and the microcanonical 
descriptions are equivalent and the model shows a continuous phase 
transition from a low-energy (temperature) ferromagnetic 
phase to a high-energy (temperature) paramagnetic phase 
for all $d\geq 3$ \cite{Joyce:inDombGreen}.

As pointed out in 1968 by H.\ E.\ Stanley, 
the free energy of a class of models described by the Hamiltonian
\begin{equation}\label{Ham_stanley_On}
{\mathbb{H}}^{(n)}=-\sum_{\langle i,j\rangle}^N 
\mathbf{T}^{(n)}_{i} \cdot \mathbf{T}^{(n)}_{j}=
-\sum_{\langle i,j\rangle}^{N} \sum_{a = 1}^n T^a_i T^a_j
\end{equation}
(with $\mathbf{T}^{(n)}_i \equiv (T_i^1,\ldots,T_i^n)$ 
and $\,|\mathbf{T}_i|^2=n\;\forall i=1,\dots,N$) 
approaches the free energy of the spherical model (\ref{Ham_sph_model}) 
in the $n\rightarrow\infty$ limit \cite{Stanley:physrev1968}. 
Moreover some ``critical properties'' of 
${\mathbb{H}}^{(n)}$, like the value of the critical temperature 
$T^{(n)}_{c}$ or the value of some critical exponents 
\cite{Stanley:prl1968}, 
appear to be monotonic functions of $n$ 
\footnote{In \cite{Stanley:prl1968} the monotonicity is explicitly shown for 
the above quantities in $d=1,2,3$ and for 
particular geometries of the lattices, i.e., spin chains, triangular 
lattices and fcc lattices. 
These results are supposed to hold also in more general cases 
but the generalization is not straightforward. 
In particular, it is not immediately clear whether the monotonicity 
is expected to hold also also for 
the energy density $\varepsilon^{(n)}_{c}$ of models 
defined by Eq.\ (\ref{H-On}) on regular cubic lattices in $d=3$.}.

The class of models described by the Hamiltonian 
in Eq.\ (\ref{Ham_stanley_On}) can be mapped onto classical 
O$(n)$ models defined by Eq.\ (\ref{H-On}), 
once the norm of the spins is properly scaled:
\begin{equation}
{\mathbb{H}}^{(n)}=-\sum_{\langle i,j\rangle}^{N} \mathbf{T}^{(n)}_{i} 
\cdot \mathbf{T}^{(n)}_{j}=
- n \sum_{\langle i,j\rangle}^{N} \mathbf{S}_{i}\cdot \mathbf{S}_{j}=  n\; H^{(n)}\, ,
\end{equation}
so that 
\begin{equation}
\lim_{n,\; N \rightarrow \infty} \frac{1}{n\;N}{\mathbb{H}}^{(n)} = 
\lim_{N \rightarrow \infty} \frac{1}{N} H^{(n)} = \lim_{N \rightarrow \infty} \frac{1}{N} H^{sph}.
\end{equation}
This implies that the thermodynamic properties of the continuous O$(n)$ 
models described by the Hamiltonian in 
Eq.\ (\ref{H-On}) converge to those of the spherical model in the $n \rightarrow \infty$ limit. 
In particular, the discrete set of critical values of the energy density: 
$\{\varepsilon^{(1)}_{c}$, $\varepsilon^{(2)}_{c}$, $\varepsilon^{(3)}_{c}$, $\varepsilon^{(4)}_{c},\dots\}$ should 
converge to $\varepsilon^{(\infty)}_{c}$ 
---that is to the critical energy density value of $H^{sph}$--- 
in the $n\rightarrow\infty$  limit. 
This means that the spherical model has to be considered an  
O$(\infty)$ model in our analysis of the critical energy 
densities. The above property hold independently 
of the spatial dimensionality $d$ of 
the lattice, hence also in the case $d=3$. 

In \cite{Binney:book,Kastner:jstat2009} an explicit expression 
for $\varepsilon^{(\infty)}_{c}$ is given: 
when adapted to our conventions in $d=3$ the result is
\begin{equation}\label{KastnerSMenergy}
\varepsilon^{(\infty)}_{c}=-3\;\; \frac{a_3}{1+a_3}\, , 
\end{equation}
where the coefficient $a_3$ is given by 
\begin{equation}\label{coefficient_a3}
a_{3}=\int_{[0,\pi]^{3}}\frac{d^{3}k}{\pi^{3}} \; 
\frac{\sum_{j=1}^{3}\cos{k_{j}}}{3-\sum_{j=1}^{3}\cos{k_{j}}}\, .
\end{equation} 
The coefficient $a_3$ is related to the Watson integral $W_3$ commonly 
used in the spherical model \cite{Joyce:inDombGreen,JoyceZucker:jphysa2001}: 
some properties of the Watson integrals are recalled in 
Appendix\ \ref{Watson}. The result 
for $a_3$ is 
\begin{equation}\label{coefficient_a3_analitico}
a_3=\frac{\sqrt{3} -1}{32 \pi^3} 
\left( \Gamma\left( \frac{1}{24} \right) 
\Gamma\left( \frac{11}{24} \right) \right)^2 -1\, ,
\end{equation}
where $\Gamma$ denotes the gamma function. 
Using (\ref{coefficient_a3_analitico}), 
the numerical value we get from Eq.\ (\ref{KastnerSMenergy}) is
\begin{equation}\label{KastnerSMenergy2}
\varepsilon^{(\infty)}_{c}=-1.0216119\dots
\end{equation}
and we shall use it 
as the critical energy density of the O$(\infty)$ model in $d=3$. 

\section{Comparison of critical energy densities}\label{Sec_NumericalTest}
The critical energy densities $\varepsilon^{(n)}_{c}$, discussed in the previous Sections for $n=1,2,3,4$ and $\infty$, 
are collected in Table \ref{table:energy_summary_results} as a function of $1/n= 1/\infty, 1/4, 1/3, 1/2$ and $1$, 
together with their derivation method.
\begin{table}[ht]
\caption{ {Critical energy densities $\varepsilon^{(n)}_{c}$ with their 
derivation method for $n=1,2,3,4$ and $n=\infty$.}}
\centering
\begin{tabular}{|c |c |c |c|}
\hline
$\frac{1}{n}$ & $\varepsilon^{(n)}_{c}$ & Derivation method \\[0.5ex]
\hline
\hline 
$\frac{1}{\infty}\equiv0$ & $-1.0216119\dots$   & Exact solution \\[0.5ex] 
$\frac{1}{4}$& $-0.99174\;\pm\;0.00002 \;\pm\;0.0006$  & FSS this work, Eq.\  (\ref{O4_energy_final_best_result})\\[0.5ex]
$\frac{1}{3}$ & $-0.989556\;\pm\;0.000010\;\pm\;0.00010$ & FSS this work, Eq.\  (\ref{H_energy_final_best_result})\\[0.5ex]
$\frac{1}{2}$ & $-0.98904\;\pm\;0.00003\;\pm\;0.0003$ & FSS this work, Eq.\  (\ref{XY_energy_best_result})\\[0.5ex]
1 & $-0.99063 \pm 0.00004$ & FSS \cite{HasenbuschPinn:jphysa1998} \\
\hline
\end{tabular}
\label{table:energy_summary_results}
\end{table}

Data in Table {\ref{table:energy_summary_results}} 
can be interpolated to obtain an estimate of $\varepsilon^{(n)}_{c}$ 
for any $n$. To make such an interpolation more reliable, 
we exploit a theoretical result by Campostrini et al.\ 
\cite{CampostriniEtAl:npb1996}. These authors performed 
an analysis of the four-point renormalized coupling 
constant in classical O$(n)$ 
models. Interestingly, an important byproduct of their study was to have an
estimate of the critical energy density $\varepsilon^{(n)}_{c}$ 
for large values of $n$, i.e., at the first order in a $1/n$ expansion. 
They found
\begin{equation}\label{CPRV}
 \varepsilon^{(n)}_{c}=\varepsilon^{(\infty)}_{c}+b_{1}\,\frac{1}{n}+O\left(\frac{1}{n^{2}}\right)\, ,
\end{equation}
and the numerical result for the coefficient $b_1$ given in \cite{CampostriniEtAl:npb1996}, once adapted to our conventions, is $b_1 = 0.21$. The accuracy of $b_1$ affects the accuracy of the interpolation, as we shall see below, hence we repeated the numerical calculation of $b_1$ increasing its precision; as reported in Appendix\ \ref{Watson}, we obtained $b_1=0.2182(8)$.
 
This result suggests an interpolation of the data in 
Table \ref{table:energy_summary_results} has to be performed: 
$\varepsilon_{c}(n)$ should be a polynomial function in $\frac{1}{n}$ 
in which the zero-order term is given 
by the critical energy density $\varepsilon^{(\infty)}_{c}$ of the spherical 
model as given in Eq.\ (\ref{KastnerSMenergy2}), 
and the coefficient of the linear term is fixed to $b_1$. 
Using these constraints and the data of Table \ref{table:energy_summary_results},  we numerically computed the interpolating 
function and found
\begin{equation}\label{polynomial-fit}
\varepsilon_{c}({n}) =  ~\varepsilon^{(\infty)}_{c} + b_1 
\,\frac{1}{n} +b_{2} \, \frac{1}{n^{2}} +b_{3}\, \frac{1}{n^{3}} 
+b_{4}\, \frac{1}{n^{4}}
\end{equation}
finding $b_2=-0.4762$, $b_3=0.3105$ and $b_4=0.0593$. 
In the interpolation procedure we did not consider 
the point $\{ 1, \varepsilon^{(1)}_{c} \}$ since our interest is in 
the comparison of $\varepsilon^{(n \geq 2)}_{c}$ and $\varepsilon^{(1)}_{c}$ in 
$\frac{1}{n}\in\left[ 0,\frac{1}{2} \right]$. Moreover, 
the function $\varepsilon_{c}(n)$ has to be computed with the lowest order polynomial function as possible. If we force 
$\varepsilon_{c}(n)$ to pass through $\{1,\varepsilon^{(1)}_{c}\}$, 
the next-order term ($b_{5}\,\frac{1}{n^{5}}$) becomes necessary although no 
useful information on $\varepsilon^{(n)}_{c}$ 
is present in the range $1/n\in[1/2,1]$.
As a further check we also performed a fit of data presented 
in Table \ref{table:energy_summary_results} 
(without the point $\{1,\varepsilon^{(1)}_{c}\}$) with a fourth-order polynomial obtaining an excellent agreement with the interpolation. 

However, the value of $b_1$ is known with a finite precision, and this affects the reliability of the numerical values of the coefficients $b_2$, $b_3$ and $b_4$. To estimate the accuracy of the coefficients of the interpolation formula we thus repeated the procedure using $b_1 = 0.2190$ and $b_1 = 0.2174$, i.e., the upper and lower bounds for $b_1$, respectively. We can summarize the results as follows: the interpolation formula for the critical energy density is given by Eq.\ \eqref{polynomial-fit} with $\varepsilon^{(\infty)}_{c} = -1.0216119\dots$, $b_1 = 0.2182(8)$, $b_2 = -0.472(7)$, $b_3 = 0.31(2)$ and $b_4 = 0.06(2)$.

\begin{figure}
\centering
\includegraphics[width=10cm]{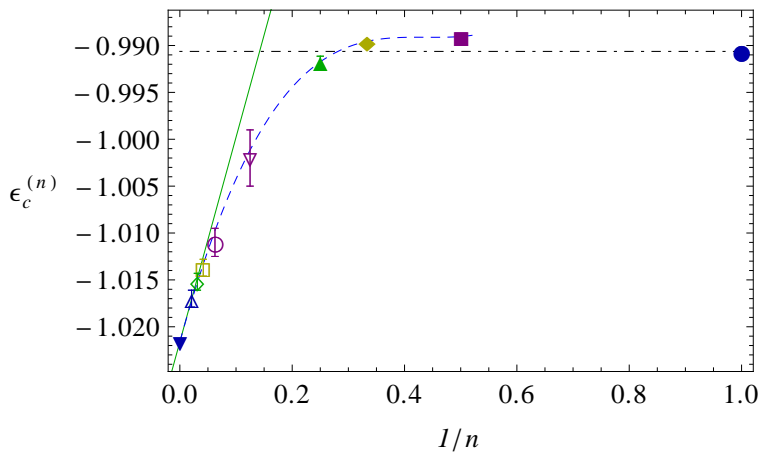} 
\caption{Critical energy densities $\varepsilon^{(n)}_{c}$ of 3-$d$ O$(n)$ models 
as a function of $1/n$: $\varepsilon^{(1)}_{c}$ (solid blue circle), $\varepsilon^{(2)}_{c}$ (solid purple square), $\varepsilon^{(3)}_{c}$ 
(solid yellow diamond), $\varepsilon^{(4)}_{c}$ (solid green up-pointing triangle) and $\varepsilon^{(\infty)}_{c}$ (solid blue down-pointing 
triangle) as given in Table \ref{table:energy_summary_results}; uncertainties are smaller than or of the same order of the symbol sizes.
The dashed blue line is the interpolating curve $\varepsilon_{c}(1/n)$ given in Eq.\ (\ref{polynomial-fit}) with the coefficients given in the text, the solid 
green line represents the $\frac{1}{n}$ expansion up to first order as given in Eq.\ (\ref{CPRV}), the horizontal dot-dashed black line is the line 
of equation $\varepsilon_{c}(n)=\varepsilon^{(1)}_{c}$. Open symbols are the values of the critical energies found by strong-coupling expansion in Ref.\ \cite{CampostriniEtAl:npb1996} for $n = 8$ (open purple down-pointing triangle), $n = 16$ (open purple circle), $n = 24$ (open yellow square), $n = 32$ (open green diamond), and $n = 48$ (open blue up-pointing triangle).}
\label{final-energy-plot}
\end{figure}

In Fig.\ \ref{final-energy-plot} we plot the following quantities: 
the interpolating curve given by Eq.\ (\ref{polynomial-fit}) with the above reported coefficients (dashed blue line), the first-order approximation as given by Eq.\ (\ref{CPRV}) (solid green line), the horizontal curve 
$\varepsilon^{(n)}_{c}=\varepsilon^{(1)}_{c}$ in correspondence of the 
critical energy density of the Ising model 
(dot-dashed black line), and, with solid symbols, the critical energy densities 
$\varepsilon^{(1)}_{c}$, $\varepsilon^{(2)}_{c}$ 
(purple square), $\varepsilon^{(3)}_{c}$, 
$\varepsilon^{(4)}_{c}$ and $\varepsilon^{(\infty)}_{c}$ (blue down-pointing triangle). 
For $1/n=1/2,1/3,1/4$ the uncertainties on the points are given by 
the systematic uncertainties shown in Table \ref{table:energy_summary_results} and are hardly visible on the plot being smaller than the symbols' size. 
Simulation data for $n$ larger than 4 are not available. We thus reported on the plot the values of $\varepsilon^{(4)}_{c}$ obtained in Ref.\ \cite{CampostriniEtAl:npb1996} with a strong-coupling expansion, using open symbols. Although these data are less accurate than simulation data they are in very good agreement with the interpolation formula.

The interpolating curve provides a practical test for 
the reliability of the approximation 
$\varepsilon^{(n)}_{c}\simeq \varepsilon^{(1)}_{c}$ discussed 
at the beginning. Indeed, assuming that Eq.\ (\ref{polynomial-fit}) 
yields good estimates of the values of $\varepsilon^{(n)}_{c}$, 
for any $n\in[2,\infty]$ the discrepancy between 
$\varepsilon^{(n)}_{c}$ and 
$\varepsilon^{(1)}_{c}$ can be easily quantified as 
$|\varepsilon_{c}(1/n)-\varepsilon^{(1)}_{c}|$. In particular: 
for $1/n\in[1,1/8)$, that is up to $n=8$, the error committed by 
replacing $\varepsilon^{(n)}_{c}$ with $\varepsilon^{(1)}_{c}$ 
is about $1\%$; for $1/n\in[1/8,1/18)$, 
that is up to $n=18$, the error is about $2\%$; 
for $1/n\in[1/18,0]$, that is up to $n=\infty$, 
the error is about $3\%$, and in any case smaller than 
$|\varepsilon^{(\infty)}_{c}-\varepsilon^{(1)}_{c}|\simeq 0.031$. We checked that the same conclusion is obtained by performing a fit of the form (\ref{polynomial-fit}) using also the data for $\varepsilon^{(n)}_{c}$ with $n=8,16,24,32,48$ reported in \cite{CampostriniEtAl:npb1996} (and of course the data of Table \ref{table:energy_summary_results}).

\section{Concluding remarks}\label{ConclusionsNumericalTest}
We have performed a numerical analysis of the $n$-dependence of 
the critical energy density of three-dimensional 
classical O$(n)$ models defined on regular cubic lattices 
and with nearest-neighbor ferromagnetic interactions: our results 
are summarized in Table \ref{table:energy_summary_results}. 
For $n=2$ and $3$, our results for the critical energy densities 
---Eqs.\ (\ref{XY_energy_I_best_result}) and (\ref{H_energy_I_best_result})--- 
improved the accuracy of the numerical estimates present in the literature.

The critical energy densities of classical O$(n)$ models with 
$n=2,3$ and $4$ have been evaluated with a finite-size 
scaling (FSS) analysis together with their statistical 
and systematic uncertainties due to the FSS procedure 
and to the uncertainty on the critical temperature, respectively; 
the systematic uncertainties turned out to be much larger 
(an order of magnitude)
than the statistical ones 
for every value of $n$. 
A possible way to further reduce these systematic uncertainties 
in future simulations would possibly be to compute the critical temperature 
$T_c^{(n)}(L)$ at size $L$ \cite{Binder:book}, vary $L$ and then proceed 
to the FSS analysis.

Interpolating the data of $\varepsilon^{(n)}_{c}$ 
for $n=2,3,4$ and $n=\infty$, a polynomial function $\varepsilon_{c}(n)$ 
has been computed to estimate the critical energy density at any $n$. 
This function exploits the knowledge of the first-order term in the 
$1/n$-expansion of the critical energy density of O$(n)$ models 
computed in \cite{CampostriniEtAl:npb1996}, and 
yields a a practical way to test the error committed 
by replacing $\varepsilon^{(n)}_{c}$ with $\varepsilon^{(1)}_{c}$ for a 
generic O$(n)$ model. The latter is less than $1\%$ if $n\in[2,8)$, 
between $1\%$ and $2\%$ if $n\in[8,18]$ and 
less then $3\%$ for all the larger $n$'s up to $n=\infty$. 

The above analysis concludes the discussion started in \cite{prl2011} 
as to the values of the critical energy densities of classical O$(n)$ 
models with ferromagnetic interactions defined on regular cubic lattices 
in $d=3$, showing that the critical energy densities of these models 
are indeed very close to each other and quantifying their differences. 
Clearly this result alone does not mean that 
the rather crude approximations on the density of states 
put forward in \cite{prl2011} are reliable. 
However, as already recalled in the Introduction, 
such approximations can be controlled and a relation similar 
to \eqref{omega_appr} can be derived for two exactly solvable models, 
the mean-field and 1-$d$ $XY$ models \cite{jstat2012}, 
and similar considerations can be effectively 
used to construct analytical or semi-analytical 
estimates of the density of states of O$(n)$ models 
that compare well with simulation data 
for $n=2$ in $d=2$ \cite{analyticalpaper}.


Finally a comment is in order on the critical energy  densities 
for three-dimensional O$(n)$ models found in this paper. 
As briefly discussed in Sec.\ \ref{numericalSphericalModel}, 
a monotonic behavior in $n$ is supposed to hold for 
some thermodynamic functions of classical O$(n)$ models defined 
on particular lattice geometries \cite{Stanley:prl1968}. 
It is unclear whether such considerations 
could be applied also to $\varepsilon^{(n)}_{c}$ of O$(n)$ models 
defined on regular cubic lattices. The interpolating function 
in Eq.\ (\ref{polynomial-fit}) is a monotonically 
increasing function of $\frac{1}{n}$ from $n =\infty$ up to $n=2$, 
but this is no longer true for $n=1$ since ---within the estimated errors--- 
it is $\varepsilon^{(1)}_{c} < \varepsilon^{(2)}_{c}$. 
Monotonicity could be restored admitting a higher value 
$\varepsilon^{(1)\prime}_{c}$ 
for $\varepsilon^{(1)}_{c}$, 
such that $\varepsilon^{(1)\prime}_{c}-\varepsilon^{(1)}_{c}\simeq10^{-3}$. 
The accuracy of the numerical value of $\varepsilon^{(1)}_{c}$ 
in Eq.\ (\ref{Ising_best_energy}) derived  
in \cite{HasenbuschPinn:jphysa1998} clearly does not 
allow such a higher value of $\varepsilon^{(1)}_{c}$. 
Hence we conclude that monotonicity fails for $n=1$, unless 
the uncertainty quoted in \cite{HasenbuschPinn:jphysa1998} 
is underestimated. However, a possible increase 
of $10^{-3}$ in $\varepsilon^{(1)}_{c}$ would neither affect the 
considerations made at the end of Sec.\ \ref{Sec_NumericalTest} 
nor the form of Eq.\ (\ref{polynomial-fit}). 

\acknowledgments
Discussions with E.\ Vicari and G.\ Gori are gratefully acknowledged.

\appendix
\section{Some properties of the Watson integrals and estimate of $b_1$}\label{Watson}
The Watson integrals appear in the 
theory of the spherical model \cite{Joyce:inDombGreen} and 
are related to the generalized Watson integrals
\begin{equation}
W(d,z)=\frac{1}{\pi^d} \, \int_0^\pi \cdots \int_0^\pi \, 
\frac{dk_1 \cdots dk_d}{1-\frac{1}{dz}
\left( \cos{k_1} + \cdots + \cos{k_d} \right)}\, .
\end{equation}
The Watson integral in dimension $d$ is defined as
\begin{equation}\label{Watson_Wd}
W_d=\frac{1}{\pi^d} \, \int_0^\pi \cdots \int_0^\pi \,
\frac{dk_1 \cdots dk_d}
{d-\left( \cos{k_1} + \cdots + \cos{k_d} \right)}\, , 
\end{equation}
so that 
\begin{equation}
d\,W_d=W(d,1)\, .
\end{equation}
Using the notation 
\begin{equation}\label{Watson_notation}
f_d(\mathbf{k}) \equiv d- \sum_{\alpha=1}^d \cos{k_\alpha}\, 
\end{equation}
with $\mathbf{k} = (k_1,\ldots,k_d)$, 
the Watson integral $W_d$ can be compactly written in the form
\begin{equation}\label{Watson_notation_W}
W_d=\int_{[0,\pi]^d} 
\, \frac{d^dk}{\pi^d} \, \frac{1}{f_d(\mathbf{k})}\, .
\end{equation}
The coefficient $a_d$ defined in Eq.\ (\ref{coefficient_a3}) for $d=3$ 
reads in dimension $d$
\begin{equation}
a_d=\int_{[0,\pi]^d} 
\, \frac{d^dk}{\pi^d} \, \frac{\sum_{\alpha=1}^d \cos{k_\alpha}}{f_d(\mathbf{k})}
\, : 
\end{equation}
$a_d$ is related to the Watson integral $W_d$ according to the relation
\begin{equation}
a_d=d\,W_d-1\, . 
\end{equation}

A major simplification in the evaluation of Watson integrals is obtained 
by using the identity \cite{Maradudin:book} 
\begin{equation}\label{identity}
\frac{1}{\lambda}=\int_0^{\infty} \, e^{-\lambda t} \, dt\, :
\end{equation}
by putting $\lambda=f_d(\mathbf{k})=d-\sum_{\alpha=1}^d \cos{k_\alpha}$ in 
Eq.\ (\ref{Watson_Wd}) and integrating over the $k_\alpha$'s one gets 
the \textit{single} integral
\begin{equation}\label{identity_Wd}
W_d=\int_0^{\infty} \, e^{-d t} \, [I_0(t)]^d \, dt\, ,
\end{equation}
where $I_0(t)=(1/\pi) \, \int_0^\pi e^{t \cos{k}} \, dk$ is a modified  
Bessel function of the first kind.

In $d=3$ it is possible to write $W_d$ in terms of the gamma function 
\cite{Watson:QJMO1939,Joyce:jphysa1972,BorweinZucker:IMA1992} as
\begin{equation}\label{identity_Wd_d3}
W_3=\frac{\sqrt{3} -1}{96 \pi^3} 
\left( \Gamma\left( \frac{1}{24} \right) 
\Gamma\left( \frac{11}{24} \right) \right)^2 \, ,
\end{equation}
from which Eq.\ (\ref{coefficient_a3_analitico}) follows.

The Watson integral in $d=3$ and its generalizations enter as well in the 
coefficients of the $1/n$ expansion 
\cite{CampostriniEtAl:npb1996,MullerRuhl:AnnPhys1986}: in particular 
the coefficient $b_1$ defined in the expression\ (\ref{CPRV}) for the 
critical energy density reads as \cite{CampostriniEtAl:npb1996}  
\begin{equation}\label{b1_app}
b_1= 2 
\left( \frac{b_1^{(a)}}{4} 
- \frac{1}{W_3} - \frac {b_1^{(b)}}{\left( W_3 \right)^2} 
\right) \, ,
\end{equation}
where the coefficients $b_1^{(a)},b_1^{(b)}$ are computed as integrals 
of the function $\Delta(\mathbf{q})$ defined as
\begin{equation}\label{Delta_app}
\frac{1}{\Delta(\mathbf{q})} = \frac{1}{8} \int_{[-\pi,\pi]^3} 
\frac{d^3k}{(2\pi)^3} \, \frac{1}
{f_3\left( \mathbf{k} \right) \, f_3\left( \mathbf{k+q} \right)} \, ,
\end{equation}
with $\mathbf{q}$ belonging to the first Brillouin zone 
($\mathbf{q} \in [\pi,\pi]^3$) and $f_3\left( \mathbf{k} \right)=
3-\sum_{\alpha=x,y,z} \cos{k_\alpha}$. We observe that using twice the identity 
(\ref{identity}) one can formally reduce the integral in (\ref{Delta_app}) 
to a double integral as
\begin{equation}\label{Delta_app_twice}
\frac{1}{\Delta(\mathbf{q})} = \frac{1}{8} \int_0^\infty dt_1  \, e^{-3t_1} 
\int_0^\infty dt_2  \, e^{-3t_2} \, \left( \prod_{\alpha=x,y,z} {\cal I}
\left(q_\alpha;t_1,t_2  \right) \right) \, 
\end{equation} 
[similarly to the re-writing (\ref{identity_Wd}) for $W_d$] with 
\begin{equation}\label{Delta_app_twice_I}
{\cal I} \left(q;t_1,t_2 \right) = \int_{-\pi}^{\pi} \, \frac{dk}{2\pi} \, 
e^{\,t_1\,\cos{k}+t_2\,\cos{(k+q)}}\, .
\end{equation} 
The expressions for $b_1^{(a)}$ and $b_1^{(b)}$ are respectively given by
\begin{equation}\label{b1_a}
b_1^{(a)}=\frac{1}{2} \,  \int_{[-\pi,\pi]^3} 
\frac{d^3q}{(2\pi)^3} \, \frac{\Delta(\mathbf{q})}
{f_3\left( \mathbf{q} \right) } \, 
\end{equation} 
and 
\begin{equation}\label{b1_b}
b_1^{(b)}=-\frac{1}{16} \,  \int_{[-\pi,\pi]^3} \frac{d^3q}{(2\pi)^3} \, 
\Delta(\mathbf{q}) \int_{[-\pi,\pi]^3} \frac{d^3p}{(2\pi)^3} \frac{1}
{\left( f_3\left( \mathbf{p} \right) \right)^2} \left[ 
\frac{1}{f_3\left( \mathbf{p+q} \right)} + 
\frac{1}{f_3\left( \mathbf{p-q} \right)} - 
\frac{2}{f_3\left( \mathbf{q} \right)} \right]\, .
\end{equation} 
Numerically we obtained $b_1^{(a)}=6.49628(1)$ and $b_1^{(b)}=-0.1184(1)$, 
from which $b_1=0.2182(8)$.

\bibliography{/Users/casetti/Work/Scripta/papers/bib/mybiblio,/Users/casetti/Work/Scripta/papers/bib/statmech}

\begin{thebibliography}{41}%
\makeatletter
\providecommand \@ifxundefined [1]{%
 \@ifx{#1\undefined}
}%
\providecommand \@ifnum [1]{%
 \ifnum #1\expandafter \@firstoftwo
 \else \expandafter \@secondoftwo
 \fi
}%
\providecommand \@ifx [1]{%
 \ifx #1\expandafter \@firstoftwo
 \else \expandafter \@secondoftwo
 \fi
}%
\providecommand \natexlab [1]{#1}%
\providecommand \enquote  [1]{``#1''}%
\providecommand \bibnamefont  [1]{#1}%
\providecommand \bibfnamefont [1]{#1}%
\providecommand \citenamefont [1]{#1}%
\providecommand \href@noop [0]{\@secondoftwo}%
\providecommand \href [0]{\begingroup \@sanitize@url \@href}%
\providecommand \@href[1]{\@@startlink{#1}\@@href}%
\providecommand \@@href[1]{\endgroup#1\@@endlink}%
\providecommand \@sanitize@url [0]{\catcode `\\12\catcode `\$12\catcode
  `\&12\catcode `\#12\catcode `\^12\catcode `\_12\catcode `\%12\relax}%
\providecommand \@@startlink[1]{}%
\providecommand \@@endlink[0]{}%
\providecommand \url  [0]{\begingroup\@sanitize@url \@url }%
\providecommand \@url [1]{\endgroup\@href {#1}{\urlprefix }}%
\providecommand \urlprefix  [0]{URL }%
\providecommand \Eprint [0]{\href }%
\providecommand \doibase [0]{http://dx.doi.org/}%
\providecommand \selectlanguage [0]{\@gobble}%
\providecommand \bibinfo  [0]{\@secondoftwo}%
\providecommand \bibfield  [0]{\@secondoftwo}%
\providecommand \translation [1]{[#1]}%
\providecommand \BibitemOpen [0]{}%
\providecommand \bibitemStop [0]{}%
\providecommand \bibitemNoStop [0]{.\EOS\space}%
\providecommand \EOS [0]{\spacefactor3000\relax}%
\providecommand \BibitemShut  [1]{\csname bibitem#1\endcsname}%
\let\auto@bib@innerbib\@empty
\bibitem [{\citenamefont {Casetti}\ \emph {et~al.}(2011)\citenamefont
  {Casetti}, \citenamefont {Nardini},\ and\ \citenamefont
  {Nerattini}}]{prl2011}%
  \BibitemOpen
  \bibfield  {author} {\bibinfo {author} {\bibfnamefont {L.}~\bibnamefont
  {Casetti}}, \bibinfo {author} {\bibfnamefont {C.}~\bibnamefont {Nardini}}, \
  and\ \bibinfo {author} {\bibfnamefont {R.}~\bibnamefont {Nerattini}},\ }\href
  {\doibase 10.1103/PhysRevLett.106.057208} {\bibfield  {journal} {\bibinfo
  {journal} {Phys. Rev. Lett.}\ }\textbf {\bibinfo {volume} {106}},\ \bibinfo
  {pages} {057208} (\bibinfo {year} {2011})}\BibitemShut {NoStop}%
\bibitem [{\citenamefont {Nardini}\ \emph {et~al.}(2012)\citenamefont
  {Nardini}, \citenamefont {Nerattini},\ and\ \citenamefont
  {Casetti}}]{jstat2012}%
  \BibitemOpen
  \bibfield  {author} {\bibinfo {author} {\bibfnamefont {C.}~\bibnamefont
  {Nardini}}, \bibinfo {author} {\bibfnamefont {R.}~\bibnamefont {Nerattini}},
  \ and\ \bibinfo {author} {\bibfnamefont {L.}~\bibnamefont {Casetti}},\ }\href
  {\doibase 10.1088/1742-5468/2012/02/P02007} {\bibfield  {journal} {\bibinfo
  {journal} {Journal of Statistical Mechanics: Theory and Experiment}\ }\textbf
  {\bibinfo {volume} {2012}},\ \bibinfo {pages} {P02007} (\bibinfo {year}
  {2012})}\BibitemShut {NoStop}%
\bibitem [{\citenamefont {Nardini}\ \emph {et~al.}(2013)\citenamefont
  {Nardini}, \citenamefont {Nerattini},\ and\ \citenamefont
  {Casetti}}]{analyticalpaper}%
  \BibitemOpen
  \bibfield  {author} {\bibinfo {author} {\bibfnamefont {C.}~\bibnamefont
  {Nardini}}, \bibinfo {author} {\bibfnamefont {R.}~\bibnamefont {Nerattini}},
  \ and\ \bibinfo {author} {\bibfnamefont {L.}~\bibnamefont {Casetti}},\
  }\href@noop {} {\  (\bibinfo {year} {2013})},\ \Eprint
  {http://arxiv.org/abs/arXiv:1312.5223} {arXiv:1312.5223} \BibitemShut
  {NoStop}%
\bibitem [{\citenamefont {Wales}(2004)}]{Wales:book}%
  \BibitemOpen
  \bibfield  {author} {\bibinfo {author} {\bibfnamefont {D.~J.}\ \bibnamefont
  {Wales}},\ }\href@noop {} {\emph {\bibinfo {title} {Energy Landscapes}}}\
  (\bibinfo  {publisher} {Cambridge University Press},\ \bibinfo {address}
  {Cambridge},\ \bibinfo {year} {2004})\BibitemShut {NoStop}%
\bibitem [{Note1()}]{Note1}%
  \BibitemOpen
  \bibinfo {note} {The relation \protect \textup {\hbox {\mathsurround \z@
  \protect \normalfont (\ignorespaces \ref {omega_appr}\unskip \@@italiccorr
  )}} cannot be exact, at least in the form proposed in \protect \cite
  {prl2011}, because it would imply wrong ---and $n$-independent--- values of
  the critical exponent $\alpha $. Nevertheless Eq.\ (\ref {omega_appr}) yields
  the correct sign of $\alpha $, that is, correctly predicts a cusp in the
  specific heat at criticality and not a divergence: see Refs.\ \cite {prl2011}
  and especially \cite {jstat2012} for a more complete discussion on the
  problem.}\BibitemShut {Stop}%
\bibitem [{\citenamefont {Campa}\ \emph {et~al.}(2003)\citenamefont {Campa},
  \citenamefont {Giansanti},\ and\ \citenamefont
  {Moroni}}]{CampaGiansantiMoroni:jpa2003}%
  \BibitemOpen
  \bibfield  {author} {\bibinfo {author} {\bibfnamefont {A.}~\bibnamefont
  {Campa}}, \bibinfo {author} {\bibfnamefont {A.}~\bibnamefont {Giansanti}}, \
  and\ \bibinfo {author} {\bibfnamefont {D.}~\bibnamefont {Moroni}},\ }\href
  {\doibase 10.1088/0305-4470/36/25/301} {\bibfield  {journal} {\bibinfo
  {journal} {Journal of Physics A: Mathematical and General}\ }\textbf
  {\bibinfo {volume} {36}},\ \bibinfo {pages} {6897} (\bibinfo {year}
  {2003})}\BibitemShut {NoStop}%
\bibitem [{\citenamefont {Archambault}\ \emph {et~al.}(1997)\citenamefont
  {Archambault}, \citenamefont {Bramwell},\ and\ \citenamefont
  {Holdsworth}}]{Archambault_etal:jpa1997}%
  \BibitemOpen
  \bibfield  {author} {\bibinfo {author} {\bibfnamefont {P.}~\bibnamefont
  {Archambault}}, \bibinfo {author} {\bibfnamefont {S.~T.}\ \bibnamefont
  {Bramwell}}, \ and\ \bibinfo {author} {\bibfnamefont {P.~C.~W.}\ \bibnamefont
  {Holdsworth}},\ }\href {\doibase 10.1088/0305-4470/30/24/005} {\bibfield
  {journal} {\bibinfo  {journal} {Journal of Physics A: Mathematical and
  General}\ }\textbf {\bibinfo {volume} {30}},\ \bibinfo {pages} {8363}
  (\bibinfo {year} {1997})}\BibitemShut {NoStop}%
\bibitem [{\citenamefont {de~Souza}\ and\ \citenamefont
  {Brady~Moreira}(1993)}]{BradyMoreira:prb1993}%
  \BibitemOpen
  \bibfield  {author} {\bibinfo {author} {\bibfnamefont {A.~J.~F.}\
  \bibnamefont {de~Souza}}\ and\ \bibinfo {author} {\bibfnamefont {F.~G.}\
  \bibnamefont {Brady~Moreira}},\ }\href {\doibase 10.1103/PhysRevB.48.9586}
  {\bibfield  {journal} {\bibinfo  {journal} {Phys. Rev. B}\ }\textbf {\bibinfo
  {volume} {48}},\ \bibinfo {pages} {9586} (\bibinfo {year}
  {1993})}\BibitemShut {NoStop}%
\bibitem [{\citenamefont {Gottlob}\ and\ \citenamefont
  {Hasenbusch}(1993)}]{GottlobHasenbusch:physicaa1993}%
  \BibitemOpen
  \bibfield  {author} {\bibinfo {author} {\bibfnamefont {A.~P.}\ \bibnamefont
  {Gottlob}}\ and\ \bibinfo {author} {\bibfnamefont {M.}~\bibnamefont
  {Hasenbusch}},\ }\href {\doibase 10.1016/0378-4371(93)90131-M} {\bibfield
  {journal} {\bibinfo  {journal} {Physica A: Statistical Mechanics and its
  Applications}\ }\textbf {\bibinfo {volume} {201}},\ \bibinfo {pages} {593 }
  (\bibinfo {year} {1993})}\BibitemShut {NoStop}%
\bibitem [{\citenamefont {Brown}\ and\ \citenamefont
  {Ciftan}(2006)}]{BrownCiftan:prb2006}%
  \BibitemOpen
  \bibfield  {author} {\bibinfo {author} {\bibfnamefont {R.~G.}\ \bibnamefont
  {Brown}}\ and\ \bibinfo {author} {\bibfnamefont {M.}~\bibnamefont {Ciftan}},\
  }\href {\doibase 10.1103/PhysRevB.74.224413} {\bibfield  {journal} {\bibinfo
  {journal} {Phys. Rev. B}\ }\textbf {\bibinfo {volume} {74}},\ \bibinfo
  {pages} {224413} (\bibinfo {year} {2006})}\BibitemShut {NoStop}%
\bibitem [{\citenamefont {Engels}\ and\ \citenamefont
  {Karsch}(2012)}]{EngelsKarsch:prd2012}%
  \BibitemOpen
  \bibfield  {author} {\bibinfo {author} {\bibfnamefont {J.}~\bibnamefont
  {Engels}}\ and\ \bibinfo {author} {\bibfnamefont {F.}~\bibnamefont
  {Karsch}},\ }\href {\doibase 10.1103/PhysRevD.85.094506} {\bibfield
  {journal} {\bibinfo  {journal} {Phys. Rev. D}\ }\textbf {\bibinfo {volume}
  {85}},\ \bibinfo {pages} {094506} (\bibinfo {year} {2012})}\BibitemShut
  {NoStop}%
\bibitem [{\citenamefont {Stanley}(1968{\natexlab{a}})}]{Stanley:physrev1968}%
  \BibitemOpen
  \bibfield  {author} {\bibinfo {author} {\bibfnamefont {H.~E.}\ \bibnamefont
  {Stanley}},\ }\href {\doibase 10.1103/PhysRev.176.718} {\bibfield  {journal}
  {\bibinfo  {journal} {Phys. Rev.}\ }\textbf {\bibinfo {volume} {176}},\
  \bibinfo {pages} {718} (\bibinfo {year} {1968}{\natexlab{a}})}\BibitemShut
  {NoStop}%
\bibitem [{\citenamefont {Campostrini}\ \emph {et~al.}(1996)\citenamefont
  {Campostrini}, \citenamefont {Pelissetto}, \citenamefont {Rossi},\ and\
  \citenamefont {Vicari}}]{CampostriniEtAl:npb1996}%
  \BibitemOpen
  \bibfield  {author} {\bibinfo {author} {\bibfnamefont {M.}~\bibnamefont
  {Campostrini}}, \bibinfo {author} {\bibfnamefont {A.}~\bibnamefont
  {Pelissetto}}, \bibinfo {author} {\bibfnamefont {P.}~\bibnamefont {Rossi}}, \
  and\ \bibinfo {author} {\bibfnamefont {E.}~\bibnamefont {Vicari}},\ }\href
  {\doibase http://dx.doi.org/10.1016/0550-3213(95)00569-2} {\bibfield
  {journal} {\bibinfo  {journal} {Nuclear Physics B}\ }\textbf {\bibinfo
  {volume} {459}},\ \bibinfo {pages} {207 } (\bibinfo {year}
  {1996})}\BibitemShut {NoStop}%
\bibitem [{\citenamefont {Hasenbusch}\ and\ \citenamefont
  {Pinn}(1998)}]{HasenbuschPinn:jphysa1998}%
  \BibitemOpen
  \bibfield  {author} {\bibinfo {author} {\bibfnamefont {M.}~\bibnamefont
  {Hasenbusch}}\ and\ \bibinfo {author} {\bibfnamefont {K.}~\bibnamefont
  {Pinn}},\ }\href {\doibase 10.1088/0305-4470/31/29/007} {\bibfield  {journal}
  {\bibinfo  {journal} {Journal of Physics A: Mathematical and General}\
  }\textbf {\bibinfo {volume} {31}},\ \bibinfo {pages} {6157} (\bibinfo {year}
  {1998})}\BibitemShut {NoStop}%
\bibitem [{Note2()}]{Note2}%
  \BibitemOpen
  \bibinfo {note} {But in the case $n\rightarrow \infty $, that will be
  discussed in Sec.\ \ref {numericalSphericalModel}.}\BibitemShut {Stop}%
\bibitem [{ALP()}]{ALPS}%
  \BibitemOpen
  \href@noop {} {\bibinfo  {journal} {{Web page:}
  \texttt{http://alps.comp-phys.org/}}\ }\BibitemShut {NoStop}%
\bibitem [{\citenamefont {Fisher}(1974)}]{Fisher:rmp1974}%
  \BibitemOpen
\bibfield  {journal} {  }\bibfield  {author} {\bibinfo {author} {\bibfnamefont
  {M.~E.}\ \bibnamefont {Fisher}},\ }\href {\doibase 10.1103/RevModPhys.46.597}
  {\bibfield  {journal} {\bibinfo  {journal} {Rev. Mod. Phys.}\ }\textbf
  {\bibinfo {volume} {46}},\ \bibinfo {pages} {597} (\bibinfo {year}
  {1974})}\BibitemShut {NoStop}%
\bibitem [{\citenamefont {Br\`{e}zin}(1982)}]{Brezin:jphys1982}%
  \BibitemOpen
  \bibfield  {author} {\bibinfo {author} {\bibfnamefont {E.}~\bibnamefont
  {Br\`{e}zin}},\ }\href@noop {} {\bibfield  {journal} {\bibinfo  {journal} {J.
  Physique}\ }\textbf {\bibinfo {volume} {43}},\ \bibinfo {pages} {15}
  (\bibinfo {year} {1982})}\BibitemShut {NoStop}%
\bibitem [{\citenamefont {Stanley}(1999)}]{Stanley:rmp1999}%
  \BibitemOpen
  \bibfield  {author} {\bibinfo {author} {\bibfnamefont {H.~E.}\ \bibnamefont
  {Stanley}},\ }\href {\doibase 10.1103/RevModPhys.71.S358} {\bibfield
  {journal} {\bibinfo  {journal} {Rev. Mod. Phys.}\ }\textbf {\bibinfo {volume}
  {71}},\ \bibinfo {pages} {S358} (\bibinfo {year} {1999})}\BibitemShut
  {NoStop}%
\bibitem [{\citenamefont {Schultka}\ and\ \citenamefont
  {Manousakis}(1995)}]{SchultkaManousakis:prb1995}%
  \BibitemOpen
  \bibfield  {author} {\bibinfo {author} {\bibfnamefont {N.}~\bibnamefont
  {Schultka}}\ and\ \bibinfo {author} {\bibfnamefont {E.}~\bibnamefont
  {Manousakis}},\ }\href {\doibase 10.1103/PhysRevB.52.7528} {\bibfield
  {journal} {\bibinfo  {journal} {Phys. Rev. B}\ }\textbf {\bibinfo {volume}
  {52}},\ \bibinfo {pages} {7528} (\bibinfo {year} {1995})}\BibitemShut
  {NoStop}%
\bibitem [{Note3()}]{Note3}%
  \BibitemOpen
  \bibinfo {note} {For $c^{(n)}_{c}$ only the statistical error $\varDelta
  c^{(n),stat}_{c}$ will be computed since this quantity is only used for the
  computation of $\varDelta \varepsilon ^{(n),syst}_{c}$.}\BibitemShut {Stop}%
\bibitem [{Note4()}]{Note4}%
  \BibitemOpen
  \bibinfo {note} {Notice that Eqs.\ (\ref {energy_FSS}) and Eq.\ (\ref
  {energy_FSS_out}) hold for $T=T^{(n)}_{c}$ - however, since $\protect \frac
  {\varDelta T^{(n)}_{c}}{T^{(n)}_{c}}\sim 10^{-5}$ for the models considered,
  we assume Eq.\ (\ref {energy_FSS_out}) valid in the whole range $T\in \left
  [T^{(n)}_{c}-\varDelta T^{(n)}_{c}, T^{(n)}_{c}+\varDelta T^{(n)}_{c}\right
  ]$.}\BibitemShut {Stop}%
\bibitem [{\citenamefont {Talapov}\ and\ \citenamefont
  {Bl\"{o}te}(1996)}]{TalapovBlote:jphysa1996}%
  \BibitemOpen
  \bibfield  {author} {\bibinfo {author} {\bibfnamefont {A.~L.}\ \bibnamefont
  {Talapov}}\ and\ \bibinfo {author} {\bibfnamefont {H.~W.~J.}\ \bibnamefont
  {Bl\"{o}te}},\ }\href {\doibase 10.1088/0305-4470/29/17/042} {\bibfield
  {journal} {\bibinfo  {journal} {Journal of Physics A: Mathematical and
  General}\ }\textbf {\bibinfo {volume} {29}},\ \bibinfo {pages} {5727}
  (\bibinfo {year} {1996})}\BibitemShut {NoStop}%
\bibitem [{\citenamefont {Landau}\ and\ \citenamefont
  {Binder}(2009)}]{Binder:book}%
  \BibitemOpen
  \bibfield  {author} {\bibinfo {author} {\bibfnamefont {D.~P.}\ \bibnamefont
  {Landau}}\ and\ \bibinfo {author} {\bibfnamefont {K.}~\bibnamefont
  {Binder}},\ }\href@noop {} {\emph {\bibinfo {title} {A Guide to Monte Carlo
  Simulations in Statistical Physics}}},\ \bibinfo {edition} {3rd}\ ed.\
  (\bibinfo  {publisher} {Cambridge University Press},\ \bibinfo {address}
  {Cambridge},\ \bibinfo {year} {2009})\BibitemShut {NoStop}%
\bibitem [{\citenamefont {Goldner}\ and\ \citenamefont
  {Ahlers}(1992)}]{GoldnerAhlers:prb1992}%
  \BibitemOpen
  \bibfield  {author} {\bibinfo {author} {\bibfnamefont {L.~S.}\ \bibnamefont
  {Goldner}}\ and\ \bibinfo {author} {\bibfnamefont {G.}~\bibnamefont
  {Ahlers}},\ }\href {\doibase 10.1103/PhysRevB.45.13129} {\bibfield  {journal}
  {\bibinfo  {journal} {Phys. Rev. B}\ }\textbf {\bibinfo {volume} {45}},\
  \bibinfo {pages} {13129} (\bibinfo {year} {1992})}\BibitemShut {NoStop}%
\bibitem [{\citenamefont {Hasenbusch}\ and\ \citenamefont
  {T\"or\"ok}(1999)}]{HasenbuschTorok:jphysa1999}%
  \BibitemOpen
  \bibfield  {author} {\bibinfo {author} {\bibfnamefont {M.}~\bibnamefont
  {Hasenbusch}}\ and\ \bibinfo {author} {\bibfnamefont {T.}~\bibnamefont
  {T\"or\"ok}},\ }\href {\doibase 10.1088/0305-4470/32/36/301} {\bibfield
  {journal} {\bibinfo  {journal} {Journal of Physics A: Mathematical and
  General}\ }\textbf {\bibinfo {volume} {32}},\ \bibinfo {pages} {6361}
  (\bibinfo {year} {1999})}\BibitemShut {NoStop}%
\bibitem [{\citenamefont {Le~Guillou}\ and\ \citenamefont
  {Zinn-Justin}(1980)}]{LeGuillouZinnJustin:prb1980}%
  \BibitemOpen
  \bibfield  {author} {\bibinfo {author} {\bibfnamefont {J.~C.}\ \bibnamefont
  {Le~Guillou}}\ and\ \bibinfo {author} {\bibfnamefont {J.}~\bibnamefont
  {Zinn-Justin}},\ }\href {\doibase 10.1103/PhysRevB.21.3976} {\bibfield
  {journal} {\bibinfo  {journal} {Phys. Rev. B}\ }\textbf {\bibinfo {volume}
  {21}},\ \bibinfo {pages} {3976} (\bibinfo {year} {1980})}\BibitemShut
  {NoStop}%
\bibitem [{\citenamefont {Holm}\ and\ \citenamefont
  {Janke}(1994)}]{HolmJanke:jphysa1994}%
  \BibitemOpen
  \bibfield  {author} {\bibinfo {author} {\bibfnamefont {C.}~\bibnamefont
  {Holm}}\ and\ \bibinfo {author} {\bibfnamefont {W.}~\bibnamefont {Janke}},\
  }\href {\doibase 10.1088/0305-4470/27/7/030} {\bibfield  {journal} {\bibinfo
  {journal} {Journal of Physics A: Mathematical and General}\ }\textbf
  {\bibinfo {volume} {27}},\ \bibinfo {pages} {2553} (\bibinfo {year}
  {1994})}\BibitemShut {NoStop}%
\bibitem [{\citenamefont {Kanaya}\ and\ \citenamefont
  {Kaya}(1995)}]{KanayaKaya:prd1995}%
  \BibitemOpen
  \bibfield  {author} {\bibinfo {author} {\bibfnamefont {K.}~\bibnamefont
  {Kanaya}}\ and\ \bibinfo {author} {\bibfnamefont {S.}~\bibnamefont {Kaya}},\
  }\href {\doibase 10.1103/PhysRevD.51.2404} {\bibfield  {journal} {\bibinfo
  {journal} {Phys. Rev. D}\ }\textbf {\bibinfo {volume} {51}},\ \bibinfo
  {pages} {2404} (\bibinfo {year} {1995})}\BibitemShut {NoStop}%
\bibitem [{\citenamefont {Berlin}\ and\ \citenamefont
  {Kac}(1952)}]{BerlinKac:physrev1952}%
  \BibitemOpen
  \bibfield  {author} {\bibinfo {author} {\bibfnamefont {T.~H.}\ \bibnamefont
  {Berlin}}\ and\ \bibinfo {author} {\bibfnamefont {M.}~\bibnamefont {Kac}},\
  }\href {\doibase 10.1103/PhysRev.86.821} {\bibfield  {journal} {\bibinfo
  {journal} {Phys. Rev.}\ }\textbf {\bibinfo {volume} {86}},\ \bibinfo {pages}
  {821} (\bibinfo {year} {1952})}\BibitemShut {NoStop}%
\bibitem [{\citenamefont {Binney}\ \emph {et~al.}(1992)\citenamefont {Binney},
  \citenamefont {Dowrick}, \citenamefont {Fisher},\ and\ \citenamefont
  {Newman}}]{Binney:book}%
  \BibitemOpen
  \bibfield  {author} {\bibinfo {author} {\bibfnamefont {J.~J.}\ \bibnamefont
  {Binney}}, \bibinfo {author} {\bibfnamefont {N.~J.}\ \bibnamefont {Dowrick}},
  \bibinfo {author} {\bibfnamefont {A.~J.}\ \bibnamefont {Fisher}}, \ and\
  \bibinfo {author} {\bibfnamefont {M.}~\bibnamefont {Newman}},\ }\href@noop {}
  {\emph {\bibinfo {title} {The Theory of {C}ritical {P}henomena: {A}n
  {I}ntroduction to the {R}enormalization {G}roup}}}\ (\bibinfo  {publisher}
  {Oxford University Press},\ \bibinfo {address} {New York},\ \bibinfo {year}
  {1992})\BibitemShut {NoStop}%
\bibitem [{\citenamefont {Kastner}(2009)}]{Kastner:jstat2009}%
  \BibitemOpen
  \bibfield  {author} {\bibinfo {author} {\bibfnamefont {M.}~\bibnamefont
  {Kastner}},\ }\href {\doibase 10.1088/1742-5468/2009/12/P12007} {\bibfield
  {journal} {\bibinfo  {journal} {Journal of Statistical Mechanics: Theory and
  Experiment}\ }\textbf {\bibinfo {volume} {2009}},\ \bibinfo {pages} {P12007}
  (\bibinfo {year} {2009})}\BibitemShut {NoStop}%
\bibitem [{\citenamefont {Joyce}(1972{\natexlab{a}})}]{Joyce:inDombGreen}%
  \BibitemOpen
  \bibfield  {author} {\bibinfo {author} {\bibfnamefont {G.~S.}\ \bibnamefont
  {Joyce}},\ }in\ \href@noop {} {\emph {\bibinfo {booktitle} {{P}hase
  {T}ransitions and {C}ritical {P}henomena}}},\ Vol.~\bibinfo {volume} {2},\
  \bibinfo {editor} {edited by\ \bibinfo {editor} {\bibfnamefont
  {C.}~\bibnamefont {Domb}}\ and\ \bibinfo {editor} {\bibfnamefont {M.~S.}\
  \bibnamefont {Green}}}\ (\bibinfo  {publisher} {Academic Press},\ \bibinfo
  {year} {1972})\BibitemShut {NoStop}%
\bibitem [{\citenamefont {Stanley}(1968{\natexlab{b}})}]{Stanley:prl1968}%
  \BibitemOpen
  \bibfield  {author} {\bibinfo {author} {\bibfnamefont {H.~E.}\ \bibnamefont
  {Stanley}},\ }\href {\doibase 10.1103/PhysRevLett.20.589} {\bibfield
  {journal} {\bibinfo  {journal} {Phys. Rev. Lett.}\ }\textbf {\bibinfo
  {volume} {20}},\ \bibinfo {pages} {589} (\bibinfo {year}
  {1968}{\natexlab{b}})}\BibitemShut {NoStop}%
\bibitem [{Note5()}]{Note5}%
  \BibitemOpen
  \bibinfo {note} {In \cite {Stanley:prl1968} the monotonicity is explicitly
  shown for the above quantities in $d=1,2,3$ and for particular geometries of
  the lattices, i.e., spin chains, triangular lattices and fcc lattices. These
  results are supposed to hold also in more general cases but the
  generalization is not straightforward. In particular, it is not immediately
  clear whether the monotonicity is expected to hold also also for the energy
  density $\varepsilon ^{(n)}_{c}$ of models defined by Eq.\ (\ref {H-On}) on
  regular cubic lattices in $d=3$.}\BibitemShut {Stop}%
\bibitem [{\citenamefont {Joyce}\ and\ \citenamefont
  {Zucker}(2001)}]{JoyceZucker:jphysa2001}%
  \BibitemOpen
  \bibfield  {author} {\bibinfo {author} {\bibfnamefont {G.~S.}\ \bibnamefont
  {Joyce}}\ and\ \bibinfo {author} {\bibfnamefont {I.~J.}\ \bibnamefont
  {Zucker}},\ }\href {\doibase 10.1088/0305-4470/34/36/314} {\bibfield
  {journal} {\bibinfo  {journal} {Journal of Physics A: Mathematical and
  General}\ }\textbf {\bibinfo {volume} {34}},\ \bibinfo {pages} {7349}
  (\bibinfo {year} {2001})}\BibitemShut {NoStop}%
\bibitem [{\citenamefont {Maradudin}\ \emph {et~al.}(1960)\citenamefont
  {Maradudin}, \citenamefont {Montroll}, \citenamefont {Weiss}, \citenamefont
  {Herman},\ and\ \citenamefont {Milnes}}]{Maradudin:book}%
  \BibitemOpen
  \bibfield  {author} {\bibinfo {author} {\bibfnamefont {A.~A.}\ \bibnamefont
  {Maradudin}}, \bibinfo {author} {\bibfnamefont {E.~W.}\ \bibnamefont
  {Montroll}}, \bibinfo {author} {\bibfnamefont {G.~H.}\ \bibnamefont {Weiss}},
  \bibinfo {author} {\bibfnamefont {R.}~\bibnamefont {Herman}}, \ and\ \bibinfo
  {author} {\bibfnamefont {H.~W.}\ \bibnamefont {Milnes}},\ }\href@noop {}
  {\emph {\bibinfo {title} {Green's Functions for Monoatomic Simple Cubic
  Lattice}}}\ (\bibinfo  {publisher} {Acad\'{e}mie Royale de Belgique},\
  \bibinfo {address} {Bruxelles},\ \bibinfo {year} {1960})\BibitemShut
  {NoStop}%
\bibitem [{\citenamefont {Watson}(1939)}]{Watson:QJMO1939}%
  \BibitemOpen
  \bibfield  {author} {\bibinfo {author} {\bibfnamefont {G.~N.}\ \bibnamefont
  {Watson}},\ }\href@noop {} {\bibfield  {journal} {\bibinfo  {journal} {Q. J.
  Math. Oxford}\ }\textbf {\bibinfo {volume} {10}},\ \bibinfo {pages} {266}
  (\bibinfo {year} {1939})}\BibitemShut {NoStop}%
\bibitem [{\citenamefont {Joyce}(1972{\natexlab{b}})}]{Joyce:jphysa1972}%
  \BibitemOpen
  \bibfield  {author} {\bibinfo {author} {\bibfnamefont {G.~S.}\ \bibnamefont
  {Joyce}},\ }\href {\doibase 10.1088/0305-4470/5/8/001} {\bibfield  {journal}
  {\bibinfo  {journal} {Journal of Physics A: Mathematical and General}\
  }\textbf {\bibinfo {volume} {5}},\ \bibinfo {pages} {L65} (\bibinfo {year}
  {1972}{\natexlab{b}})}\BibitemShut {NoStop}%
\bibitem [{\citenamefont {Borwein}\ and\ \citenamefont
  {Zucker}(1992)}]{BorweinZucker:IMA1992}%
  \BibitemOpen
  \bibfield  {author} {\bibinfo {author} {\bibfnamefont {J.~M.}\ \bibnamefont
  {Borwein}}\ and\ \bibinfo {author} {\bibfnamefont {I.~J.}\ \bibnamefont
  {Zucker}},\ }\href@noop {} {\bibfield  {journal} {\bibinfo  {journal} {IMA J.
  Numer. Anal.}\ }\textbf {\bibinfo {volume} {12}},\ \bibinfo {pages} {519}
  (\bibinfo {year} {1992})}\BibitemShut {NoStop}%
\bibitem [{\citenamefont {M\"{u}ller}\ and\ \citenamefont
  {R\"{u}hl}(1986)}]{MullerRuhl:AnnPhys1986}%
  \BibitemOpen
  \bibfield  {author} {\bibinfo {author} {\bibfnamefont {V.~F.}\ \bibnamefont
  {M\"{u}ller}}\ and\ \bibinfo {author} {\bibfnamefont {W.}~\bibnamefont
  {R\"{u}hl}},\ }\href@noop {} {\bibfield  {journal} {\bibinfo  {journal}
  {Annals of Physics}\ }\textbf {\bibinfo {volume} {168}},\ \bibinfo {pages}
  {425} (\bibinfo {year} {1986})}\BibitemShut {NoStop}%
\end{thebibliography}%

\end{document}